\documentclass[aps,pra,reprint,floatfix,superscriptaddress]{revtex4-1}
\usepackage{hyperref}
\usepackage{amsmath}
\usepackage{graphicx}
\newcommand{\Tc}{T_\mathrm{C}}
\newcommand{\muB}{\mu_\mathrm{B}}
\usepackage{bm}
\renewcommand{\vec}[1]{\bm{#1}}

\begin{document}

\title{Rare-earth/transition-metal magnets at finite temperature:
self-interaction-corrected relativistic density functional theory
in the disordered local moment picture}
\author{Christopher E. Patrick}
\email[]{c.patrick.1@warwick.ac.uk}
\author{Julie B.\ Staunton}
\affiliation{Department of Physics, University of Warwick, Coventry CV4 7AL, United Kingdom}
\date{\today}

\begin{abstract}
Atomic-scale computational modeling of technologically relevant 
permanent magnetic materials faces two key challenges.
First, a material's magnetic properties depend sensitively on temperature,
so the calculations must account for thermally induced magnetic disorder.
Second, the most widely-used permanent magnets are based on rare-earth
elements, whose highly localized 4$f$ electrons are poorly described
by standard electronic structure methods.
Here, we take two established theories, the disordered local moment
picture of thermally induced magnetic disorder and self-interaction-corrected
density functional theory,  and devise
a computational framework to overcome these challenges.
Using the new approach, we calculate
magnetic moments and Curie temperatures of the rare-earth cobalt
(RECo$_5$) family for RE=Y--Lu.
The calculations correctly reproduce the experimentally measured trends
across the series and confirm that, apart from the hypothetical compound
EuCo$_5$, SmCo$_5$ has the strongest magnetic properties at high temperature.
An order-parameter analysis demonstrates 
that varying the RE has a surprisingly strong effect on 
the Co--Co magnetic interactions
determining the Curie temperature, even when the lattice
parameters are kept fixed.
We propose the origin of this behavior is
a small contribution to the density from $f$-character electrons 
located close to the Fermi level.
\end{abstract}

\maketitle

\section{Introduction}

In solids the 15 lanthanides (atomic numbers 57--71)
usually exist in a 3+ state, with three electrons 
(two of $s$ and one of $d$ character) donated to the valence band. 
Grouping the lanthanides with Y and Sc, which behave in the same way,
forms the group of elements known as the ``rare earths'' (REs)~\cite{Elliottbook}.
The chemical variation within the REs
originates from their strongly-localized 
4$f$ spin up/down subshells, which vary
from being totally empty (Sc/Y/La, 4$f^0$) to totally filled (Lu, 4$f^{14}$).
Lying at the centre of the lanthanide block, Gd (4$f^7$)
has one completely filled and one completely 
empty spin subshell, and marks the boundary between the ``light''
(Sc--Eu) and the ``heavy'' REs (Gd--Lu).
Notable anomalous lanthanides include Ce, whose valence
varies due to the relative ease that its single 4$f$ electron can delocalize;
Pm, which is radioactively unstable; 
and Eu and Yb which, rather than having a single hole in a spin
subshell associated with the 3+ state, usually
prefer to capture an additional $4f$ electron and adopt a 2+ 
state~\cite{Elliottbook,Gschneidnerbook}.

Aside from their uses in e.g.\ catalysts, batteries and 
energy-efficient lighting~\cite{Zepf2013},
the excellent magnetic properties of Sm-Co~\cite{Strnat1967} 
and Nd-Fe-B~\cite{Sagawa1984,Croat1984} compounds
have led to REs becoming critical to many industries
as components in high-performance permanent magnets~\cite{Gutfleisch2011}.
The key principle underlying such magnets
is that while elemental transition metals (TMs) like Fe and
Co remain strongly magnetic up to very high temperatures 
($\sim$1000~K), they are relatively easy to demagnetize with external fields~\cite{Coey2011}.
Alloying the elemental TMs with the REs largely retains their good 
high-temperature properties whilst simultaneously providing a massive 
increase in the coercivity (resistance to demagnetization)~\cite{Coey2011}.
The principal microscopic mechanism driving this increased coercivity is 
the electrostatic interaction of the localized RE-4$f$ electrons
with their environment (the crystal field)~\cite{Kuzmin2008}.
The magnetic moment
associated with the RE-4$f$ electrons gains a strong directional preference,
i.e.\ magnetocrystalline anisotropy, which anchors the TM magnetism along
the same axis through the RE-TM exchange interaction.
The benefits of this alloying approach can be seen in SmCo$_5$, 
whose Curie temperature ($\Tc$) of 1020~K
is comparable to pure Co (1360~K)~\cite{Buschow1977,Coey2011} but whose
magnetocrystalline anisotropy energy density
is 20 times larger~\cite{Chikazumibook}.
Indeed, over 50~years since its discovery the high-temperature 
properties of SmCo$_5$ remain hard to beat~\cite{Gutfleisch2011}.

This simple picture---that the RE provides
the coercivity and the TM provides the large magnetization and $\Tc$---is
an oversimplification, since it neglects the contribution
to the magnetization from the REs themselves.
A more coherent picture of RE-TM intermetallics~\cite{Kuzmin2008} models 
the RE as an array of isolated 3+ ions interacting with the crystal field
and an effective magnetic field originating from the RE-TM exchange interaction.
Diagonalization of the crystal-field Hamiltonian gives the RE contribution
to the magnetization and anisotropy.
The TM contribution is deduced empirically from RE-TM compounds
with a nonmagnetic RE, like Y~\cite{Tiesong1991}.

The crystal-field picture does an excellent job
of explaining the temperature dependence of
magnetic quantities which are heavily RE-dependent,
such as the low-temperature anisotropy and 
magnetization~\cite{Tiesong1991,Kuzmin2008}.
However, the phenomenological description of the TM 
limits the predictive power of the theory, especially
with regard to $\Tc$.
Since a material rapidly loses its magnetic properties
at temperatures in the vicinity of its Curie temperature,
having a high $\Tc$ is very useful for practical permanent magnets.
It is known experimentally that $\Tc$ is RE-dependent:
referring to the experimental review  of 
Ref.~\citenum{Buschow1977}, 
SmCo$_5$ has the highest $\Tc$ of the compounds that form
stoichiometrically as RECo$_5$
(1020~K), slightly higher than GdCo$_5$ (1014~K).
Meanwhile for the RE$_2$Co$_7$ and RE$_2$Co$_{17}$ 
series of magnets it is RE=Gd which has the highest $\Tc$ (771/1218~K)
with RE=Sm lower (713/1195~K).
The fact that Gd has the largest spin moment of the REs
might suggest some correlation of this quantity with
$\Tc$, but the RE$_2$Co$_7$ series provides the counterexamples 
of RE=Dy and Ho, whose nominal spin moments are larger than Sm 
but whose $\Tc$ is smaller (640~and~647~K)~\cite{Buschow1977}.
Magnetostructural effects could also play a role,
with the RE modifying the lattice constants and thus
the magnetic interactions~\cite{AndreevHMM}.
However it is by no means clear how these and other
effects might combine to influence $\Tc$.

A predictive, first-principles theory of the $\Tc$ of RE-TM magnets 
could provide insight into the physical
processes governing the high-temparature performance of these magnets,
and suggest strategies for further optimization.
However, such a theory is currently missing.
Density-functional theory (DFT)~\cite{Kohn1965}
provides a practical framework to perform
first-principles studies of RE-TM magnets, but 
is faced with the challenge of 
describing
with sufficient accuracy
(i) the finite-temperature
disorder of the magnetic moments and (ii) the complex
interactions between the localized RE-4$f$ electrons
and their itinerant counterparts.

\begin{figure}
\includegraphics[width=50mm]{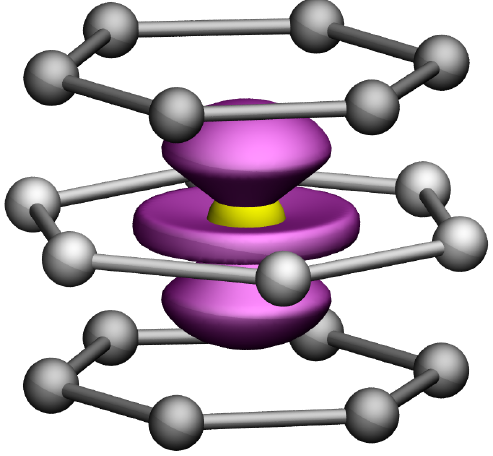}%
\caption{Schematic of the RECo$_5$ crystal structure,
showing the RE (yellow) and  Co sites
(gray).
The 2$c$ sites (referred to as Co$_\mathrm{I}$ in the text)
are in-plane with the RE, and the 3$g$ sites (Co$_\mathrm{II}$)
lie above and below.
An isosurface plot representing the Sm-4$f$ charge density,
obtained as the sum of the squared spherical harmonics with $l=3,m=3,2,...,-1$
is also shown. \label{fig.RECo5}}
\end{figure}
In this work we introduce a theory which attacks these 
two problems directly.
Finite temperature effects are modeled within
the disordered local moment 
(DLM)~\cite{Gyorffy1985,Staunton2006} picture,
which is reviewed in Section~\ref{sec.DFTDLM}.
Meanwhile the problematic RE-$4f$ electrons are 
treated within DFT using
the local self-interaction correction (LSIC)~\cite{Lueders2005}.
Previous modeling of REs within this framework has been limited
to Gd~\cite{MendiveTapia2017,Patrick2017,Petit2015,Hughes2007},
but the developments described in Section~\ref{sec.Dirac} 
now allow investigation of the entire RE series  
for the same computational cost.
We use the new theory to study the RECo$_5$ family 
of magnets (Fig.~\ref{fig.RECo5}), exploring the evolution of magnetism 
from 0~K (Sec.~\ref{sec.zeroK}) to $\Tc$ (Sec.~\ref{sec.finiteT}).
We conclude with our analysis of why, as is observed
experimentally, the calculations find SmCo$_5$ to have
the highest $\Tc$ of the RECo$_5$ magnets (Sec.~\ref{sec.discussion}).

\section{The DFT-DLM approach}

\label{sec.DFTDLM}

At finite temperature, the functional properties of all materials are modified
to some extent due to the thermal population of excited vibrational states,
e.g.\ thermal expansion or increased electrical 
resistivity~\cite{Baroni2001,Giustino2017}.
However, independent of lattice vibrations, the magnetic properties of a 
material are extremely sensitive to temperature.
The disordered local moment (DLM) picture of magnetism
provides a conceptual basis to understand this temperature variation~\cite{Gyorffy1985}.
Here the material is modeled as an array of 
microscopic magnetic moments (e.g.\ one associated with each atom), 
of fixed magnitude but variable orientation.
This picture of local moments makes no assumption that the electrons 
themselves are localized; for example,  the 3$d$ electrons responsible
for magnetism in Fe, the prototypical DLM metal, are completely itinerant~\cite{Staunton1994}.
Rather, the spin-spin correlation between electrons
near atomic sites can be strong enough to establish magnetically-polarized
regions which exist
for much longer timescales than those associated with electron motion~\cite{Gyorffy1985}.
These are the local moments.

A DLM magnetic microstate is specified by the
orientations $\{\vec{\hat{e}_i}\} = \{\vec{\hat{e}_1},...,\vec{\hat{e}_N}\}$
of the $N$ local moments.
The grand potential energy $\Omega(\{\vec{\hat{e}_i}\})$ is a function
of these local moment orientations, and the (classical) statistical
mechanics of the system is determined by the partition function
\begin{equation}
Z = \int d\vec{\hat{e}_1}d\vec{\hat{e}_2}...d\vec{\hat{e}_N} \exp\left[-\beta\Omega(\{\vec{\hat{e}_i}\})\right],
\end{equation}
where $1/\beta = k_B T$, and $T$ and $k_B$ are the temperature and Boltzmann constant.
Experimental measurements correspond to thermal averages over the magnetic microstates.
For instance, a magnetization measurement probes the average orientations
of the local moments,
\begin{equation}
\langle \vec{\hat{e}_j} \rangle_T = \frac{1}{Z}  
\int \vec{\hat{e}_j}  \ d\vec{\hat{e}_1}d\vec{\hat{e}_2}...d\vec{\hat{e}_N} 
\exp\left[-\beta\Omega(\{\vec{\hat{e}_i}\})\right].
\end{equation}
The DLM paramagnetic state corresponds to each orientation averaging to zero, 
$\langle \vec{\hat{e}_j} \rangle_T = 0$, and the highest temperature at which
$\langle \vec{\hat{e}_j} \rangle_T \neq 0$ corresponds to the Curie temperature $\Tc$.

In principle, DFT provides
a pathway to a first-principles DLM theory through the possibility of evaluating
the grand potential energy $\Omega(\{\vec{\hat{e}_i}\})$, although finding a sufficiently
accurate approximation for the exact
exchange-correlation functional remains an ongoing and formidable challenge.\cite{Cohen2008,Illas2004}
Specifically, $\Omega(\{\vec{\hat{e}_i}\})$ could be obtained from constrained
DFT calculations, with the applied constraints forcing
the local magnetizations to point along designated local moment 
directions $\{\vec{\hat{e}_i}\})$~\cite{Gyorffy1985}.
In practice however, any direct attempt to perform statistical mechanics would
soon be faced with the problem of covering the huge phase space spanned by  $\{\vec{\hat{e}_i}\}$,
requiring an effectively infinite supercell to contain all $N$ local moments.

A popular method of circumventing this problem is to replace the 
``exact'' $\Omega$ with a model, e.g.\ a Heisenberg model based on 
pairwise interactions between local moments.
The model parameters are extracted from DFT calculations, e.g.\
from the Liechtenstein formula~\cite{Liechtenstein1987} or constrained DFT~\cite{Cao2009}.
$\Tc$ is then obtained from the simpler statistical mechanics of the model, which might
be solved through a mean field approach, the random-phase approximation or Monte 
Carlo integration~\cite{Rosengaard1997,Kashyap2003,Turek2003,Kormann2009,Fukazawa2017}.

These schemes require striking a balance between a model which is sufficiently
complex to capture the necessary magnetic interactions, yet simple enough for
the statistical mechanics problem to be tractable.
The most popular pairwise model should, as its name suggests, only apply when 
the interaction between two local moments is independent of the 
alignments of all other local moments in the system.
This picture is not particularly intuitive in a metal where 
one would expect a co-operative effect, i.e.\ magnetic interactions
being reinforced when the material is in a global ferromagnetic state 
and weakened in the paramagnetic state.
Practically, this issue leads to the question of whether one should
parametrize the pairwise model for the ferromagnetic or paramagnetic
state~\cite{Liechtenstein1987}, and what to do at intermediate 
temperatures~\cite{Kormann2014}.

The DFT-DLM theory described in~\cite{Gyorffy1985} approaches the problem
in a different way.
Instead of approximating the grand potential energy, one instead introduces
an auxilliary quantity $\Omega_0(\{\vec{\hat{e}_i}\})$ with a known functional
form,
\begin{equation}
\Omega_0(\{\vec{\hat{e}_i}\}) = - \sum_i \vec{h_i}\cdot\vec{\hat{e}_i},
\label{eq.Omega0}
\end{equation}
where the ``Weiss fields'' $\{\vec{h_i}\}$ are obtained self-consistently.
Specifically, the thermodynamic inequality
\begin{equation}
F(T) \leq F_0(T) + \langle \Omega\rangle_{0,T} - \langle \Omega_0 \rangle_{0,T}
\label{eq.ineq}
\end{equation}
provides a relation between the exact free energy $F(T)$ and the free energy
of the auxiliary system, $F_0(T) = -k_BT \ln Z_0$, with
\begin{equation}
Z_0 = \prod_i \int d\vec{\hat{e}_i} \exp[\vec{\lambda_i}\cdot\vec{\hat{e}_i}]
= \prod_i \frac{4\pi}{\lambda_i} \sinh \lambda_i
\end{equation}
and $\vec{\lambda_i} = \beta \vec{h_i}$.
Crucially the thermal averages $\langle\rangle_{0,T}$ appearing in 
the inequality~\ref{eq.ineq} 
are calculated with respect to the auxilliary system, e.g.\
\begin{equation}
\langle \Omega\rangle_{0,T} = \frac{1}{Z_0}\prod_j \int d\vec{\hat{e}_j}
\exp[\vec{\lambda_j}\cdot\vec{\hat{e}_j}] 
\Omega(\{\vec{\hat{e}_i}\}).
\label{eq.P0}
\end{equation}
The Weiss fields are chosen to minimize the right
hand side of the inequality~\ref{eq.ineq}.
Then,
\begin{equation}
\vec{h_i} = -\frac{3}{4\pi}\int d\vec{\hat{e}_i} 
\langle \Omega\rangle_{0,T}^{\vec{\hat{e}_i}} \ \vec{\hat{e}_i}.
\label{eq.Weissfield}
\end{equation}
The partial average $\langle\Omega\rangle_{0,T}^{\vec{\hat{e}_i}}$ 
integrates over all the degrees of freedom in equation~\ref{eq.P0} except the single
local moment orientation $\vec{\hat{e}_i}$.
The Weiss fields have the periodicity of the magnetic unit
cell, i.e.\ the number of distinct Weiss fields equals the number
of magnetic sublattices.

As indicated by equation~\ref{eq.Weissfield}, the Weiss fields
are temperature dependent.
The DFT-DLM estimate of $\Tc$ is the temperature at which
all the Weiss fields vanish.
Alternatively, one can introduce local order parameters,
\begin{equation}
\vec{m_i}(T) \equiv  \langle \vec{\hat{e}_i} \rangle_{0,T} = 
\vec{\hat{\lambda}_i} L(\lambda_i)
\label{eq.m}
\end{equation}
with $L(\lambda_i) = \coth(\lambda_i)- 1/\lambda_i$.
These quantities vary between 1 at zero temperature and 0 at $\Tc$.

We stress that the key quantities in the DFT-DLM theory, 
the Weiss fields $\{\vec{h_i}\}$, are calculated with
the full grand potential energy $\Omega$, without any assumption
on the nature of the underlying interactions e.g.\ pairwise, four-spin etc~\cite{MendiveTapia2017}.
Furthermore, through the averaging in equation~\ref{eq.Weissfield} the
magnitudes of the Weiss fields are indeed influenced by the 
degree of global order in the system, ensuring self consistency
between  $\{\vec{h_i}\}$ and the ``reference state'' used to calculate them.

The partial average $\langle \Omega\rangle_{0,T}^{\vec{\hat{e}_i}}$ appearing
in equation~\ref{eq.Weissfield} still presents a challenge to the most widely-used 
implementations of DFT, which solve the Kohn-Sham equations to determine
single-particle wavefunctions~\cite{Kohn1965}.
However, the Green's-function-based Korringa-Kohn-Rostoker
multiple-scattering formulation of DFT, in combination with the coherent
potential approximation (KKR-CPA)~\cite{Gyorffy1979} allows the 
partial average to be  recast as an impurity problem.
This impurity problem, which sees each local moment sitting in
an effective medium designed to mimic the averaged properties
of the disordered system, can be solved using the same KKR-CPA techniques
originally developed to tackle compositional disorder in the simulation
of alloys~\cite{Ebert2011}.
The DFT-DLM theory has undergone a number of developments
from its original formulation, and is being applied to an increasingly wide range
of magnetic systems~\cite{Staunton2004,Hughes2007,Petit2015,Patrick2017,Patrick2018}.
\
The practical steps to calculating self-consistent Weiss fields
and the key multiple-scattering equations are described in 
Refs.~\cite{Patrick2017,Matsumoto2014,Staunton2006}.

The fact that the DFT-DLM theory is rooted in KKR-CPA has both 
advantages and disadvantages.
Thermally-averaged quantities, e.g.\ spin and orbital moments, 
can be calculated relatively easily by tracing the relevant operators
with the Green's function.
The calculations include both core and valence electrons explicitly,
and the structure of the KKR-CPA equations allow for a very high
degree of numerical precision e.g.\ in evaluating integrals over
the the Brillouin zone~\cite{Bruno1997}.
However, the formalism generally involves making a shape approximation 
to the Kohn-Sham potential (here we use the atomic sphere approximation, ASA)
which, although allowing a compact angular momentum basis to be used to describe
the Green's function and scattering matrices, is not expected
to perform equally well for cubic and non-cubic crystal structures~\cite{Andersen1991}.
In addition, we note that DFT-DLM is a mean-field theory, with
the Weiss fields in equation~\ref{eq.Omega0} driving the
magnetic ordering and vice versa.
With these caveats in mind, we expect trends calculated across a series 
to be more robust than absolute values of 
specific quantities.

\section{Solving the Kohn-Sham-Dirac equation within the LSIC-LSDA}
\label{sec.Dirac}
\subsection{Relativistic DFT-DLM calculations}
\label{sec.twostep}
The large atomic number of the REs
necessitates the use of relativistic (R) DFT-DLM theory to describe 
the spin-orbit coupling inherent in
RE-TM magnets as well as mass-velocity and Darwin effects.
Practically, our RDFT-DLM calculations involve two steps.
In the first step, a self-consistent, scalar-relativistic DFT
calculation is performed for a reference magnetic state.
This reference magnetic state may be fully ordered (e.g.\
a ferromagnetic arrangement of spins) or fully disordered
(the DLM/paramagnetic state).
The output of this calculation is a set of
atom-centered potentials.
In the second step these potentials are fed into the fully-relativistic 
Kohn-Sham-Dirac (KSD) equation,
thus treating spin-orbit coupling nonperturbatively.
Combining the solutions of the KSD equation with the full KKR-CPA
machinery allows the Weiss fields and $\Tc$ to be computed.
Although not a methodological necessity~\cite{Deak2014}, 
the potentials here are kept ``frozen'' in the second step, i.e.\ the density 
derived from the Green's function of the partially-ordered system is not
used to update the potentials.

When constructing the potentials in the first step, in common with
all DFT calculations it is necessary to make an approximation for
the exchange-correlation energy.
The local-spin-density approximation (LSDA)~\cite{Kohn1965,Vosko1980} 
performs rather well in describing the magnetism of itinerant electrons, 
but struggles to described the strongly-localized 4$f$ states which characterize
REs~\cite{Perdew1981}.
Furthermore, the orbital moments of transition metals are generally smaller than
observed experimentally when calculated within the LSDA~\cite{Eriksson19901}.
As a result, it is imperative to go beyond the LSDA exchange-correlation when modeling
RE-TM magnets.

\subsection{Treating RE-4$f$ electrons}
Recent computational works performed at zero temperature have employed charge-self-consistent 
dynamical mean-field theory (DMFT)~\cite{Kotliar2006}, 
in particular using the Hubbard-$I$ approximation~\cite{Lichtenstein1998},
to calculate the magnetic moments of REs~\cite{Locht2016} and RE-TM intermetallics like 
SmCo$_5$~\cite{Granas2012,Soderlind2017,Delange2017} and NdFe$_{12}$~\cite{Delange2017}.
The simpler, ``open-core'' scheme~\cite{Brooks19912} constrains
the total spin-density of the RE-4$f$ electrons to be that 
predicted by Hund's rules~\cite{Brooks19912,Richter1991,Steinbeck20012,Soderlind2017,Fukazawa2017}.
Such calculations, which provide much important insight
into RE-TM systems, bear some resemblance to crystal-field theory in the sense
that the RE-4$f$ electrons are partitioned from the rest of the material,
with the amount of hybridization they can undergo sensitive to how the calculation
is set up~\cite{Richter1991,Delange2017}.
Alternative approaches
like  LDA/GGA+$U$~\cite{Larson2004, Larson20032,Waller2016},
the orbital polarization correction (OPC)~\cite{Soderlind2014} and
the self-interaction correction (SIC)~\cite{Lebegue2005}
modify the potential at the RE site but treat all electrons
equally, in principle allowing the RE-$4f$ states to hybridize freely~\cite{Larson2004}.
An advantage of these schemes when studying trends across the RE-TM series
is that,
beyond initial choices about how the schemes are implemented, 
the calculations require minimal user input.
Indeed the parameters entering the OPC and LDA/GGA+$U$
can be calculated from first-principles, e.g.\ the Racah parameters
calculated from wavefunctions in the OPC~\cite{Eriksson19901}, or the $U$ and $J$
energies calculated from linear response~\cite{Cococcioni2005}
or constrained random-phase approximation calculations~\cite{Karlsson2010}.

The SIC, which we employ here, aims to ensure that the exchange-correlation potential
cancels the electrostatic (Hartree) energy of a single electron interacting
with itself, which is not automatically realized in the LSDA~\cite{Perdew1981}.
While the scheme becomes more complicated in extended systems, the localized nature 
of the RE-4$f$ electrons makes them particularly
suitable for the SIC~\cite{Lebegue2005}.
Furthermore, the SIC has already been formulated
within the KKR-CPA theory as the local self-interaction 
correction (LSIC)~\cite{Lueders2005}.
Indeed the LSIC has been previously used in DFT-DLM calculations to study 
Gd~\cite{MendiveTapia2017,Patrick2017,Petit2015,Hughes2007,Patrick2018}.
However, in order to treat an arbitrary RE it is necessary to 
generalize the formalism.
Conveniently, this same formalism allows the OPC to be also incorporated
in the RDFT-DLM framework, facilitating an improved  description of the Co 
orbital moments.

\subsection{An LSIC-LSDA scheme based on Hund's rules}

The  LSIC formalism~\cite{Lueders2005} is based
on applying the self-interaction correction to individual
spin and orbital angular channels, each characterized
by the pair of quantum numbers $\sigma L$.
$\sigma$ labels spin, and
$L$ is a composite quantum number which, in principle, 
labels a member of any complete set of angular momentum 
states.
In the original LSIC implementation, these angular momentum states
have the same symmetry as the nonmagnetic crystal~\cite{Lueders2005}.
However,
since the orbital moments are largely unquenched
in the RE-TM compounds, 
here we choose $L$ to label the ``atomic''
$(l,m)$ quantum numbers associated with the complex spherical
harmonics, i.e.\ eigenfunctions of the orbital angular
momentum operator $\hat{l}_z$.
As such, states that are degenerate in the nonmagnetic crystal
may be split by the LSIC.

\begin{figure}
\includegraphics{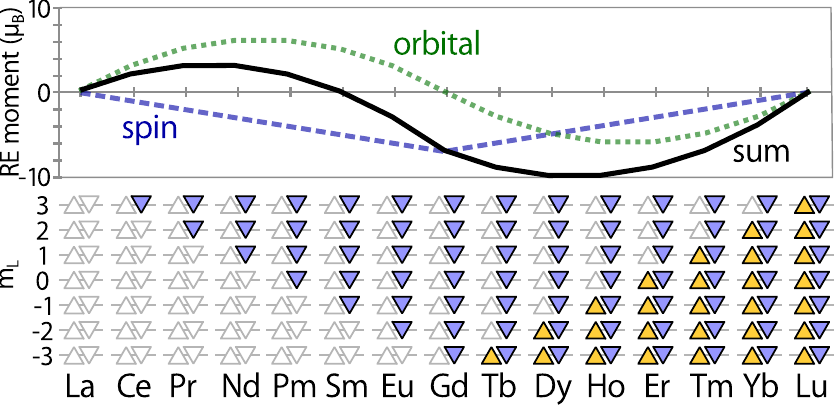}%
\caption{
Scheme to correct RE-4$f$ states based on Hund's rules.
Each triangle corresponds to a single spin and orbital angular
momentum channel i.e.\ $\sigma,l(=3),m$.
LSIC channels with $\sigma = \uparrow$ ($\downarrow$) 
are represented by yellow (blue) triangles.
We also show the moments obtained simply by adding the expectation values
of the spin and orbital operators acting on the individual corrected
states.
\label{fig.Hund}}
\end{figure}
We must also choose which spin and 
orbital angular momentum channels we should apply the LSIC to.
We propose to follow the scheme illustrated in Fig.~\ref{fig.Hund},
which is inspired by Hund's rules.
An extra LSIC channel is added for each RE-4$f$ electron,
filling up $\sigma L$ combinations of the
same spin ($\downarrow$) first with the largest available opposing $m$
(e.g.\ $m=+3$ for Ce).
After entirely filling the $\downarrow$ channel at Gd, 
we start filling the $\uparrow$ channel,
again starting
with the largest available opposing $m$ (e.g.\ $m=-3$ for Tb) 
in accordance with the single-electron
tendency of orbital and spin momenta to antialign~\cite{Chikazumibook}.
As shown in Fig.~\ref{fig.Hund}, adding up the individual spin and 
orbital angular momentum contributions associated with these filled states 
gives quantities symmetric and antisymmetric respectively about Gd.

\subsection{Including the LSIC/OPC in the KSD equation}

The LSIC scalar-relativistic calculation (the first step referred to
in Sec.~\ref{sec.twostep}) proceeds
as described in Ref.~\cite{Lueders2005}.
At the second step in the RDFT-DLM procedure (and at variance with 
previous work~\cite{MendiveTapia2017,Patrick2017,Petit2015,Hughes2007,Patrick2018})
the atom-centered potentials have a contribution which depends on angular momentum,
conveniently written as
\begin{equation}
V_\mathrm{SIC}(\vec{r}) = \sum_{L,\sigma} V_L^\sigma(r) \hat{P}_{L\sigma}.
\label{eq.Vsic}
\end{equation}
Here $V^\sigma_L(r)$ is the spherically-symmetric correction
to the potential obtained in the scalar-relativistic calculation~\cite{Lueders2005},
while $\hat{P}_{L\sigma}$ is a projection operator.
In the Pauli representation these operators are $2\times2$ matrices 
which are diagonal for spin polarization along the $z$ axis,
whose elements project out states with angular momentum
character $L$.

The angular-momentum dependent potentials result in a modified KSD equation,
\begin{eqnarray}
\left[
\vec{\tilde{\alpha}}\cdot\vec{p}c + \tilde{V}_\mathrm{SIC} + \tilde{I}(V(r) - W )\right.&& \nonumber \\
\left.+ \tilde{\beta} \left( \tilde{I}mc^2  + \tilde{\sigma}_z B_\mathrm{XC}(r) \right) \right]&&
\Psi = 0.
\label{eq.KSD}
\end{eqnarray}
Quantities with tildes are $4\times4$ matrices; $\Psi$
is a bispinor, and $W$ and $m$ the electron energy 
and rest mass.
Compared to the usual KSD equation~\cite{Strangebook}, equation~\ref{eq.KSD}
has an extra term  $\tilde{V}_\mathrm{SIC}$, simply related to
$V_\mathrm{SIC}$ in equation~\ref{eq.Vsic}:
\begin{equation}
\tilde{V}_\mathrm{SIC} = \begin{pmatrix} V_\mathrm{SIC}&0 \\ 0&V_\mathrm{SIC}\end{pmatrix}.
\end{equation}
We now follow the standard method of solving the
radial KSD equation in multiple scattering theory~\cite{Strange1984},
i.e.\ we investigate the solutions
\begin{equation}
\Psi_\nu^{m_j} (\vec{r})= \sum_{\kappa_1} \begin{pmatrix}g^{m_j}_{\kappa_1\nu} (r) |\chi_{\kappa_1}^{m_j}\rangle \\ 
if^{m_j}_{\kappa_1\nu} (r) |\chi_{-\kappa_1}^{m_j}\rangle \  \end{pmatrix}.
\label{eq.trial}
\end{equation}
The spin-angular functions $ |\chi_{\kappa_1}^{m_j}\rangle$
are superpositions of the products of Pauli spinors and spherical harmonics
weighted by Clebsch-Gordan coefficients~\cite{Strangebook}.
They are characterized by the quantum numbers $\kappa_1$ and $m_j$,
and describe the angular character of free-particle solutions 
of the KSD equation.
$\kappa_1$ is related to $j$, the sum
of spin and orbital angular momentum in the spin-angular functions,  
with $\kappa = -l - 1$ for $j = l + 1/2$ and $\kappa' = l$ 
for $j = l - 1/2$.
As indicated, we reserve the label $\kappa$ for negative
values and $\kappa'$ for positive values of $\kappa_1$.
The label $\nu$ denotes the different solutions required
to build the Green's function in scattering theory, i.e.\ solutions
with an asymptotic free-electron character 
which are regular or irregular at the origin~\cite{Strange1984}.

After inserting the trial solution~\ref{eq.trial} into equation~\ref{eq.KSD}
and performing a series of manipulations~\cite{Strangebook}, we obtain
coupled equations for the radial functions $f$ and $g$:
\begin{eqnarray}
\frac{d {f^{m_j}_{\kappa\nu}}}{dr} &=& \frac{(\kappa - 1)}{r}{f^{m_j}_{\kappa\nu}}
+ \frac{1}{\hbar c} (V-E) {g^{m_j}_{\kappa\nu}} \nonumber \\
&&+
\frac{1}{\hbar c} 
\mathcal{G}^{m_j}_+(\kappa,\kappa){g^{m_j}_{\kappa\nu} }
+
\frac{1}{\hbar c} 
\mathcal{G}^{m_j}_+(\kappa,\kappa'){g^{m_j}_{\kappa'\nu} }
\nonumber
\\
\frac{d {g^{m_j}_{\kappa\nu}}}{dr} &=& -\frac{(\kappa + 1)}{r} {g^{m_j}_{\kappa\nu}}
+ \frac{1}{\hbar c} (E-V+2mc^2) {f^{m_j}_{\kappa\nu}} \nonumber \\
&&+
\frac{1}{\hbar c} 
\mathcal{G}^{m_j}_-(-\kappa,-\kappa){f^{m_j}_{\kappa\nu}}.
\label{eq.coupled}
\end{eqnarray}
Here, $E = W - mc^2$. 
The differential equations for $f^{m_j}_{\kappa'\nu}$ and 
$g^{m_j}_{\kappa'\nu}$ are obtained from equations~\ref{eq.coupled}
simply by interchanging $\kappa$ and $\kappa'$.
Crucially, compared to previous calculations which only
included $B_\mathrm{XC}$, the basic structure of the coupled
equations~\ref{eq.coupled} is unchanged by the addition of 
$V_\mathrm{SIC}$.
The difference is in the coupling functions,
\begin{equation}
\mathcal{G}^{m_j}_\pm(\kappa_1,\kappa_2) = \langle\chi_{\kappa_1}^{m_j}|
(\sigma_z B_\mathrm{XC} \pm V_\mathrm{SIC} )
|\chi_{\kappa_2}^{m_j}\rangle.
\end{equation}
$B_\mathrm{XC}(r)$  is now augmented by a linear combination of
the LSIC potentials $V_{(l,m)\sigma}(r)$ weighted by Clebsch-Gordan
coefficients.
We give the explicit form of these coupling functions in the appendix~\ref{app.coupling},
but here just show an example of 
$\mathcal{G}^{m_j}_\pm(\kappa,\kappa)$ with $\kappa=-4$, $m_j=1/2$:
\begin{equation}
\mathcal{G}^{1/2}_\pm(-4,-4) = \frac{1}{7} B_\mathrm{XC} \pm
\left[ \frac{4}{7}V_{(3,0)}^\uparrow + \frac{3}{7}V_{(3,1)}^\downarrow\right].
\label{eq.mixing}
\end{equation}
We see that the coupling functions mix occupied, SI-corrected channels
with unoccupied, non-SI-corrected channels, as discussed more in Appendix~\ref{app.mixing}.

It should be noted that, when deriving the coupled equations~\ref{eq.coupled},
additional coupling functions of the form
$\mathcal{G}^{m_j}_\pm(-\kappa,\kappa+1)$ are introduced by both $\sigma_z$
and $V_\mathrm{SIC}$.
Follow previous work~\cite{Strange1984} we neglect these terms, which would otherwise
result in an infinite ladder of couplings between orbital angular momenta $l$, 
$l\pm2$, $l\pm4$ etc.~\cite{Ebertbook}.

The coupled equations~\ref{eq.coupled}, containing the appropriately
weighted LSIC potentials, are solved numerically to give the scattering matrices 
and regular and irregular contributions to the Green's function.
From these quantities the entire RDFT-DLM computational machinery~\cite{Matsumoto2014} 
can be applied without further modification.

The OPC enters the KSD equation in exactly the same
way as the LSIC.
This is most easily seen by writing the OPC analogy of equation~\ref{eq.Vsic} as~\cite{Steinbeck20012,Ebertbook}
\begin{equation}
V_\mathrm{OPC}(\vec{\hat{r}}) = \sum_{l=2}\sum_{m,\sigma} -B_{l\sigma} m  \langle\hat{l}_z\rangle_\sigma \hat{P}_{lm\sigma}.
\end{equation}
where $B_{l\sigma}$ is a Racah parameter, and $ \langle\hat{l}_z\rangle_\sigma$ 
is the spin-resolved expectation value for the relevant atom 
(we have anticipated applying the OPC to the $d$ channel).
Thus the OPC can be considered a special case of the LSIC where the potential
is independent of $r$, entering
$\mathcal{G}^{m_j}_\pm(\kappa_1,\kappa_2)$ weighted by the coefficients in Appendix~\ref{app.coupling}.
We stress that, since they only modify the coupling functions,
the computational cost of including the LSIC or OPC is negligible.

\subsection{Technical details}

\begin{table}
\begin{ruledtabular}
\begin{tabular}{lcccc} 
           &$a$       & $c$      & $r_\mathrm{ASA}$ & Ref.  \\
\hline
YCo$_5$    & 4.94     & 3.98     & 1.83/1.39/1.42 &\cite{AndreevHMM} \\
LaCo$_5$   & 5.11     & 3.97     & 1.91/1.40/1.44 &\cite{Buschow1977} \\
CeCo$_5$   & 4.93     & 4.01     & 1.83/1.39/1.42 &\cite{AndreevHMM} \\
PrCo$_5$   & 5.01     & 3.99     & 1.86/1.40/1.43 &\cite{AndreevHMM} \\
NdCo$_5$   & 5.01     & 3.98     & 1.86/1.40/1.43 &\cite{AndreevHMM} \\
SmCo$_5$   & 4.97     & 3.98     & 1.85/1.39/1.42 &\cite{AndreevHMM} \\
GdCo$_5$   & 4.96     & 3.97     & 1.85/1.39/1.42 &\cite{AndreevHMM} \\
TbCo$_5$   & 4.94     & 3.97     & 1.84/1.39/1.42 &\cite{AndreevHMM} \\
DyCo$_5$   & 4.91     & 3.98     & 1.82/1.38/1.41 &\cite{AndreevHMM} \\
HoCo$_5$   & 4.91     & 3.97     & 1.82/1.38/1.41 &\cite{AndreevHMM} \\
ErCo$_5$   & 4.87     & 4.00     & 1.81/1.38/1.41 &\cite{Buschow1977} \\
TmCo$_5$   & 4.86     & 4.02     & 1.81/1.38/1.41 &\cite{Buschow1977} \\
\hline
\end{tabular}
\end{ruledtabular}
\caption{
Experimental lattice constants, taken from 
Refs.~\cite{AndreevHMM,Buschow1977}.
The ASA radii for the three non-equivalent sites 
(RE/Co$_\mathrm{2c}$/Co$_\mathrm{3g}$) are also given.
All units are \AA.
\label{tab.lattice}}
\end{table}
We generate the atomic-centered potentials in the fully-ordered
(zero temperature) state
in self-consistent scalar-relativistic LSIC-LSDA calculations~\cite{Vosko1980,Lueders2005}
within the ASA, as implemented
in the \textsc{hutsepot} code~\cite{Daene2009}.
Angular momentum expansions were truncated at $l_\mathrm{max}=3$,
and the full Brillouin zone sampled on a 20$\times$20$\times$20 grid
with state occupancies determined by a Fermi-Dirac distribution
with an electronic temperature of 400~K.
The calculations were performed using experimental
lattice constants~\cite{AndreevHMM, Buschow1977},
which are listed in Table~\ref{tab.lattice} together with
the ASA radii for the three nonequivalent sites in the RECo$_5$
structure.
We used the same relations between ASA radii as in our
previous work on YCo$_5$ and GdCo$_5$~\cite{Patrick2017}.

For the RDFT-DLM calculations, apart from the inclusion
of the LSIC described above we used the same computational
setup (angular mesh, energy contour, electronic temperature)
as in~\cite{Patrick2017}, including an adaptive
sampling of the Brillouin zone~\cite{Bruno1997}.
We applied the LSIC to the RE-$4f$ electrons and the OPC
to the Co-$3d$ electrons.
The magnitude of the OPC was determined iteratively 
at 0~K with the magnetization aligned along the $c$ axis,
updating $\langle\hat{l}_z\rangle_\sigma$ at each iteration
to self-consistency.
The Racah parameters were calculated scalar-relativistically.
The same OPC was used for all temperatures,
consistent with the frozen-potential approach.

\section{Zero-temperature calculations}
\label{sec.zeroK}

\subsection{RECo$_5$ moments}

We first use the RDFT-DLM formalism to calculate the magnetic 
moments of the RECo$_5$ series at zero temperature.
To illustrate the trend across the lanthanide block
we consider all members of the RE=Y--Lu series,
including the non-forming RE = Pm, Eu, Yb and Lu compounds.
Here, we fix the lattice parameters to those
of GdCo$_5$; using the appropriate experimental RECo$_5$ 
lattice parameters (where available)
produces very similar zero-temperature moments (Appendix~\ref{app.explatt}).
For Ce, Pr and Nd we also performed calculations
without applying the LSIC (i.e.\ treating the f-electrons as itinerant).

\begin{figure}
\includegraphics{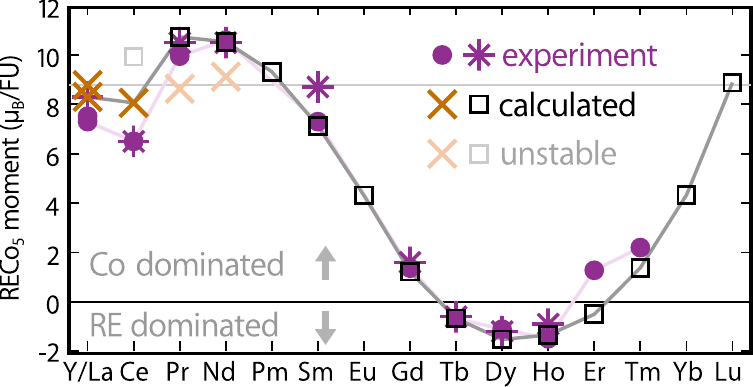}%
\caption{
Magnetic moments calculated at zero temperature with (squares) 
and without (crosses) the LSIC applied,
compared to experimental values reported in Refs.~\cite{Buschow1977} (circles)
and~\cite{AndreevHMM} (stars).
Faint symbols were calculated to be energetically unfavorable.
The gray horizontal line at 8.78$\muB$ corresponds to the calculated YCo$_5$ moment.
 \label{fig.zeroTmoms}}
\end{figure}

In Fig.~\ref{fig.zeroTmoms} we show the calculated RECo$_5$ 
moments and compare them to low-temperature experiments~\cite{Buschow1977,AndreevHMM}.
Here the Co moments are aligned along the $c$-axis, which defines
the positive direction.
A negative moment in Fig.~\ref{fig.zeroTmoms} therefore means that the
RE contribution to the magnetization is larger than that from the Co moments 
and is pointing in the opposite direction (RE-dominated).
Usually, experimental measurements (e.g.\ on powdered samples) only measure 
the absolute magnetization.
However, as we show below,
RECo$_5$ compounds which are RE-dominated
at 0~K show a compensation point (minimum) in their magnetization versus temperature 
curves, and Ref.~\cite{Buschow1977} reports compensation temperatures for
Tb, Dy and HoCo$_5$.
Accordingly we plot the experimental moments of these
three compounds with minus signs.

Considering the calculations without the LSIC first (crosses in Fig.~\ref{fig.zeroTmoms})
we see relatively small variation for different REs, with the
moments fluctuating around the YCo$_5$ value (shown as the gray
horizontal line).
We do observe a reduction in moment on moving from YCo$_5$ to LaCo$_5$,
despite both elements having an empty 4$f$ shell.
In fact, the moment of YCo$_5$ is much closer to that of LuCo$_5$,
whose 4$f$ shell is totally filled.
This behavior echoes that of quantities like melting points,
electronegativities and ionization energies, which follow
more naturally a Sc-Y-Lu series compared to Sc-Y-La~\cite{Jensen2015}.

Applying the LSIC (squares in Fig.~\ref{fig.zeroTmoms}) 
has a dramatic effect on the magnetization, for instance increasing
the moment of CeCo$_5$ by 2~$\muB$/formula unit (FU).
Now a strong variation with RE is observed, with PrCo$_5$/DyCo$_5$ achieving
the largest TM/RE-dominated moments respectively.
The transition from TM to RE-dominated magnetization occurs between Gd and Tb,
and back to TM-dominated between Er and Tm.

In order to decide whether the calculations with or without
the LSIC should be used to describe a given RE,
we examine the total energies calculated at the scalar-relativistic level
omitting spin-orbit coupling effects.
This approach follows e.g.\ Refs.~\cite{Lueders2005}~and~\cite{Strange1999},
where the comparison of SIC total energies was used to determine
the volume triggering the $\alpha\rightarrow\gamma$ transition in Ce or
the valency of the rare earths and their sulphides.
We find that
applying the LSIC to PrCo$_5$ and NdCo$_5$ lowers the total energy, 
i.e.\ it is energetically favorable.
Indeed for heavier REs the non-LSIC
calculations become difficult to converge.
However, applying the LSIC to CeCo$_5$ \emph{increases} the scalar-relativistic
total energy, indicating that the single
Ce-4$f$ electron would prefer to be delocalized in this compound.
Using this total energy as our criterion, we do not apply the LSIC to CeCo$_5$.
Indeed the picture of the itinerant Ce-4$f$ electron has already
been established in previous theoretical work~\cite{Nordstrom1990}.
Other non-energetically-favorable calculations are shown in Fig.~\ref{fig.zeroTmoms}
as faint symbols.

The variation in RECo$_5$ moment calculated with the LSIC 
largely follows the simple picture
presented in Fig.~\ref{fig.Hund}.
In general the antiferromagnetic RE-TM exchange interaction causes
the RE spin moments to point in the opposite direction to the 
Co moments~\cite{Brooks1989},
but whether or not the \emph{total} RE moment aligns parallel
or antiparallel depends on the sign and magnitude
of the orbital contribution~\cite{Nesbitt1962}.
The lightest REs have large orbital components pointing opposite to their spin
which leads to parallel alignment of the total moments,
whereas the spin and orbital moments of the heavy REs always reinforce each other
to give antiparallel alignment.

\subsection{Decomposition of RECo$_5$ moments}
\label{sec.decomp}
\begin{table*}
\begin{ruledtabular}
\begin{tabular}{lcccccc}
          &RE moment           & Scalar rel.          & Co moment          & Total moment/FU& Exp.~\cite{Buschow1977} &Exp.~\cite{AndreevHMM}\\
          &(spin/orbital/total)& spin ($f$/$spd$)     &(spin/orbital/total)&               &       &     \\
\hline
YCo$_5$ & -0.31/0.04/-0.28 & -0.31 (0.00/-0.31) & 7.54/1.25/8.78 & 8.50 & 7.52 & 8.3 \\
LaCo$_5$ & -0.30/0.04/-0.26 & -0.30 (-0.04/-0.25) & 7.11/1.19/8.30 & 8.04 & 7.3 & --- \\
CeCo$_5$ & -0.92/0.51/-0.41 & -0.86 (-0.57/-0.29) & 7.07/1.40/8.47 & 8.06 & 6.5 & 6.5 \\
CeCo$_5$*& -1.37/2.97/1.60 & -1.37 (-1.07/-0.30) & 7.19/1.14/8.33 & 9.93 & 6.5 & 6.5 \\
PrCo$_5$ & -2.46/4.88/2.42 & -2.47 (-2.13/-0.34) & 7.25/1.06/8.31 & 10.73 & 9.95 & 10.5 \\
NdCo$_5$ & -3.56/5.74/2.18 & -3.58 (-3.22/-0.37) & 7.33/1.02/8.35 & 10.53 & 10.6 & 10.5 \\
PmCo$_5$ & -4.63/5.60/0.97 & -4.71 (-4.32/-0.39) & 7.38/0.97/8.35 & 9.32 & --- & --- \\
SmCo$_5$ & -5.63/4.55/-1.08 & -5.82 (-5.41/-0.40) & 7.36/0.85/8.21 & 7.13 & 7.3 & 8.7 \\
EuCo$_5$ & -6.60/2.60/-4.01 & -6.90 (-6.48/-0.42) & 7.36/0.95/8.32 & 4.31 & --- & --- \\
GdCo$_5$ & -7.50/0.03/-7.47 & -7.49 (-7.00/-0.49) & 7.43/1.27/8.70 & 1.23 & 1.37 & 1.6 \\
TbCo$_5$ & -6.42/-2.96/-9.38 & -6.41 (-5.98/-0.44) & 7.44/1.28/8.72 & -0.67 & -0.68 & -0.6 \\
DyCo$_5$ & -5.33/-4.93/-10.26 & -5.32 (-4.93/-0.39) & 7.46/1.28/8.75 & -1.52 & -1.1 & -1.2 \\
HoCo$_5$ & -4.26/-5.88/-10.14 & -4.20 (-3.86/-0.34) & 7.51/1.29/8.80 & -1.34 & -1.49 & -0.9 \\
ErCo$_5$ & -3.28/-5.89/-9.17 & -3.09 (-2.78/-0.31) & 7.40/1.27/8.67 & -0.50 & 1.28 & --- \\
TmCo$_5$ & -2.27/-4.92/-7.19 & -2.00 (-1.71/-0.29) & 7.32/1.25/8.57 & 1.38 & 2.2 & --- \\
YbCo$_5$ & -1.26/-2.95/-4.22 & -0.92 (-0.65/-0.27) & 7.30/1.24/8.53 & 4.32 & --- & --- \\
LuCo$_5$ & -0.29/0.04/-0.25 & -0.30 (-0.03/-0.27) & 7.59/1.29/8.88 & 8.63 & --- & --- \\
\end{tabular}
\end{ruledtabular}
\caption{Decomposition of zero-temperature moments. All quantities are in $\muB$.
For comparison we include the calculations for CeCo$_5$ with the LSIC applied (*)
even though it is energetically unfavorable.
\label{tab.moments}
}
\end{table*}

In Table~\ref{tab.moments} we resolve the calculated moments
into spin and orbital contributions from the RE and TM.
We also give the spin moments calculated at the scalar-relativistic
level, which are further resolved into contributions
of different angular momentum ($f$ or $spd$) character.

Concentrating first on the RE contribution to the magnetization,
we see that the spin moments roughly track the expected spin of the 
LSIC channels, peaking at Gd.
The scalar-relativistic decomposition shows the spin moments have 
an $spd$ component which increases from 0.25$\muB$ for La to 0.49$\muB$ for Gd.
However, the $f$ components of the spin moment are not simply integers.
Based on the simple picture of Fig.~\ref{fig.Hund} this observation
is surprising, since we would expect each localized RE-$4f$ electron 
to contribute $\pm$1$\muB$ to the magnetization.
Instead, we see that for each additional LSIC channel the change in $f$ 
components is closer to $\pm$1.1$\muB$, until the elements with filled spin
subshells (GdCo$_5$ and LuCo$_5$) are reached.
This behavior indicates that the nominally unoccupied RE-4$f$ states,
which do not have the LSIC applied,
are affecting the calculated properties.

The RE orbital moments also follow the general trend of Fig.~\ref{fig.Hund}, 
but are better described by $\mu_o=(2-g_J)J$, where $g_J$ is the Land\'e factor~\cite{Richter1998}.
This textbook expression is obtained by projecting the orbital moment onto the
total angular momentum direction, which is valid for strong spin-orbit coupling.
It is therefore natural to ask whether the spin RE moments should
in fact be described by $\mu_s=2(g_J-1)J$, which is the corresponding projection
for spin~\cite{Richter1998}.
However, in our calculations the principal interaction affecting the 
spin moments is the scalar-relativistic exchange, which can be confirmed
by noting the close agreement between the RDFT-DLM and scalar-relativistic
spin moments in Table~\ref{tab.moments}.
Therefore, the spin-orbit interaction plays a relatively minor role
in determining the spin moment and the considerations leading  to 
$\mu_s$ do not apply.
We note that this situation is qualitatively different to the
open-core scheme~\cite{Brooks19912}, which fixes
the RE spin moments to $\mu_s$.

Now considering the TM contribution to the magnetization,
the most striking feature in Table~\ref{tab.moments} is the
different behavior of the light and heavy RECo$_5$ compounds.
The Co moments exhibit relatively small variations for
the heavy REs except for LuCo$_5$ which, as already noted,
behaves similarly to YCo$_5$.
However the variations for the light REs are much larger.
Moving from La to Eu, the Co spin and orbital moments
increase and decrease respectively, and in general
the total Co moments are smaller than for the heavy RECo$_5$
compounds.
As we discuss in Sec.~\ref{sec.finiteT}, a qualitative
difference in light and heavy RECo$_5$ behavior is also observed
in $\Tc$.

\subsection{Comparison to experiment}
\label{sec.mom_exp}

When comparing to experiment, it is important to note that
there is a sizeable scatter in the published data.
We have taken experimental low-temperature moments from the review
articles of Refs.~\cite{Buschow1977}~and~\cite{AndreevHMM}
which agree reasonably well with each other except
for YCo$_5$ and SmCo$_5$, which deviate by approximately
1$\muB$.
Also, we note that the RE = Tb--Tm compounds
do not form with exact RECo$_5$ stoichiometry.
Instead, due to defects where the RE is substituted 
with pairs (dumbbells) of Co atoms~\cite{Kumar1988},
the compounds become increasingly Co-rich.
For example, the actual stoichiometry of the RE=Tm compound 
reported in Ref.~\cite{Buschow1977} is TmCo$_6$.

With these limitations in mind, the calculations compare reasonably
well to experiment in Fig.~\ref{fig.zeroTmoms}.
Certainly a number of qualitative features are reproduced,
e.g.\ a drop in moment from Y to La, a large increase from Ce to Pr,
and RE-dominated magnetization for Tb--Ho.

For the special case of CeCo$_5$,
we note that the energetically-unstable LSIC calculation
gives a moment which is in qualitative disagreement
with the experimental trend.
Interestingly however, whilst the LSDA+OPC calculations are closer
to experiment they
still overestimate the CeCo$_5$ moment.
Not including the OPC on the Co atoms rather improves
the agreement (Ref.~\cite{Nordstrom1990} and Appendix~\ref{app.explatt}),
suggesting that,  (like for the LSIC), there might be a criterion
based on energetics to decide whether or not the OPC should
be applied.

Apart from the cases of ErCo$_5$ and TmCo$_5$ where the experiments
are Co-rich, the remaining compound where the discrepancy
between calculations and experiment is quite large is SmCo$_5$,
specifically compared to the value of 8.7$\muB$/FU in 
Ref.~\cite{AndreevHMM}.
Interestingly, a recent neutron diffraction experiment reported
even larger local moments in SmCo$_5$, which add up to
give a resultant magnetization in excess
of 12$\muB$/FU~\cite{Kohlmann2018}.
Studies employing DMFT and open-core calculations have reported
smaller Sm total moments of approximately -0.3$\muB$, which
would bring the total SmCo$_5$ moment closer to 
8$\muB$/FU~\cite{Granas2012,Soderlind2017,Delange2017}.
Earlier GGA+$U$ calculations found a much larger total
moment of 9.9$\muB$/FU due to a ferromagnetic alignment
of Sm and Co spins.
The scatter in theoretical and experimental data hints at
the richness of the physics of SmCo$_5$ which, as we show
next, is also seen in $\Tc$.

\section{Finite-temperature calculations}
\label{sec.finiteT}
\subsection{Magnetization vs.\ temperature curves}

\begin{figure}
\includegraphics{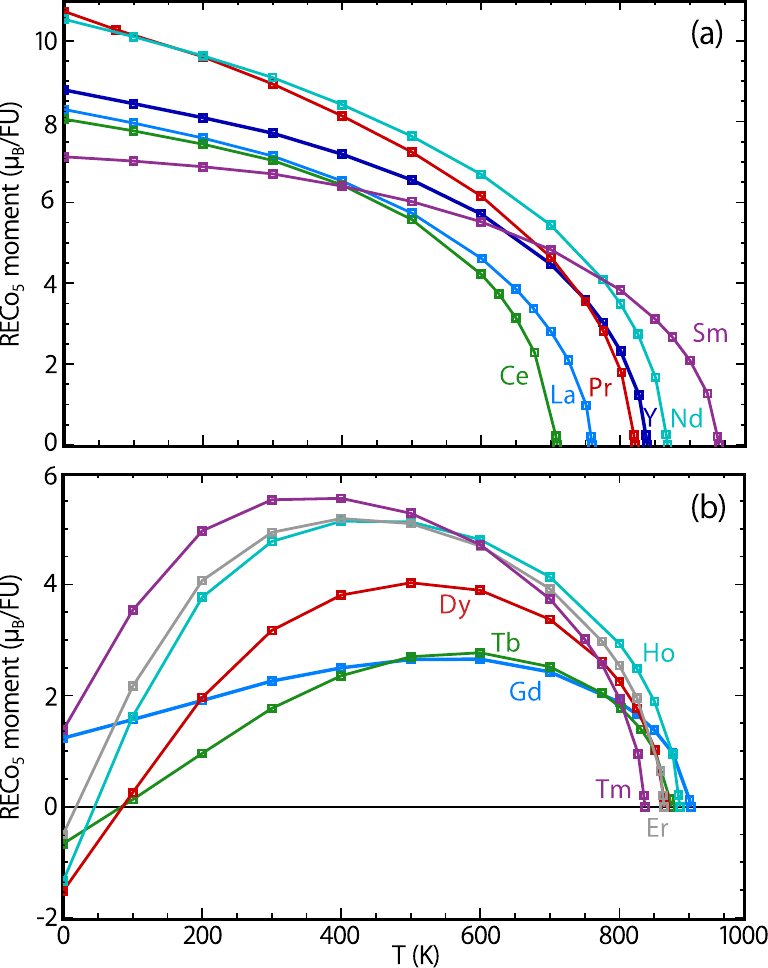}%
\caption{
Magnetization per formula unit calculated at
different temperatures for the (a) light
and (b) heavy RECo$_5$ compounds.
Calculations were performed at the GdCo$_5$ lattice parameters.
\label{fig.MvT}}
\end{figure}

We now include local moment 
disorder within the RDFT-DLM picture.
Figures~\ref{fig.MvT}(a)~and~(b) show the magnetization versus temperature ($M$v$T$) 
curves calculated for light and heavy RECo$_5$ compounds.
These calculations were performed at the GdCo$_5$ lattice constants (Table~\ref{tab.lattice}).
The light REs show behaviour associated with ferromagnets, i.e.\
a monotonic decrease in magnetization with increasing temperature.
By contrast the heavy RECo$_5$ compounds have magnetizations
which initially become more positive (TM-dominated) as the temperature
increases, before reducing at higher temperatures.
As we have shown previously by comparing YCo$_5$ and GdCo$_5$~\cite{Patrick2017}
this contrasting behaviour is due to the RE moments disordering more quickly with
temperature compared to the antiferromagnetically-aligned Co sublattice.
As a result, the strong negative contribution to the total magnetization from
the heavy RE diminishes quickly, leaving the positive Co magnetization.

In the case that the zero-temperature magnetization is RE-dominated, there
is a compensation temperature at which the strongly-disordered RE magnetization
cancels the weakly-disordered Co magnetization.
Our calculated compensation temperatures are 84~K (TbCo$_5$),
85~K (DyCo$_5$), 45~K (HoCo$_5$) and 19~K (ErCo$_5$).
Ref.~\cite{Buschow1977} reports experimental compensation
temperatures of 110~K (TbCo$_5$), 123~K (DyCo$_5$) and 71~K (HoCo$_5$).

We note that the calculated $M$v$T$ curves have finite slopes
at $T=0$~K, while experimentally-measured curves tend to be flat~\cite{Patrick2017}.
The origin of this discrepancy is the classical statistical mechanics used 
in the DLM picture (equation~\ref{eq.Omega0}), which does not give
an energy barrier between the zero-temperature arrangement of local moments
and an excited state where the moments have undergone infinitesimal rotations.

\subsection{RE order parameters}
\begin{figure}
\includegraphics{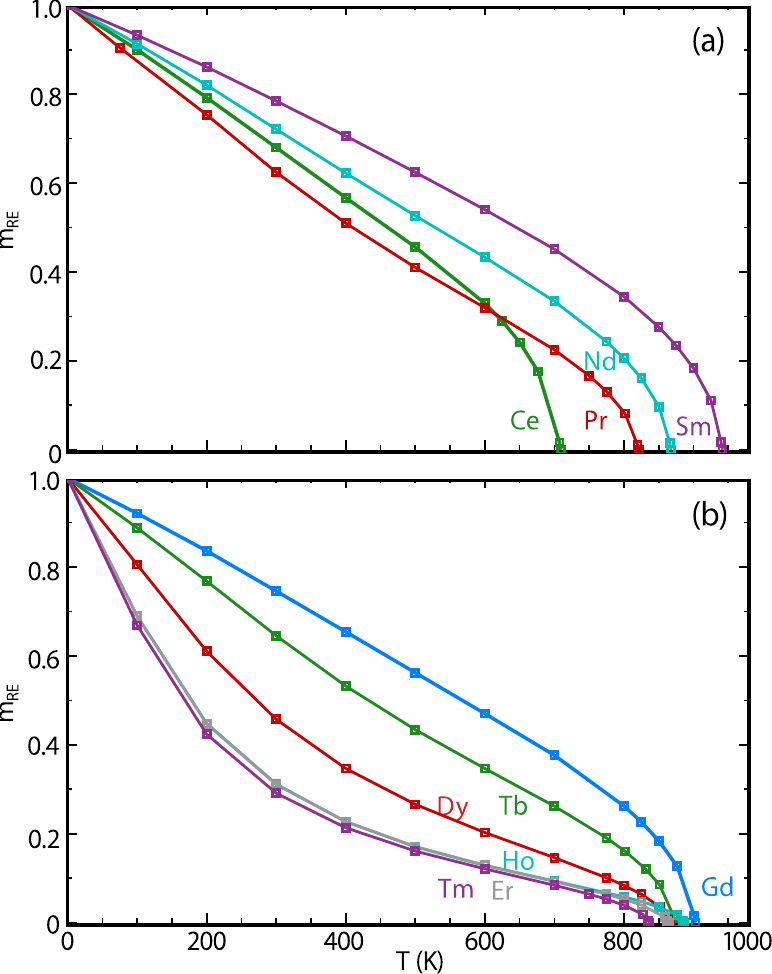}%
\caption{
RE order parameters $m_\mathrm{RE}$ (equation~\ref{eq.m}) from the calculations of Fig.~\ref{fig.MvT}
on the light (a) and heavy (b) RECo$_5$ compounds.
\label{fig.orderparams}}
\end{figure}

In order to analyse the RE contribution to the magnetization in more detail,
in Figs.~\ref{fig.orderparams}(a)~and(b) we plot the temperature evolution of the
RE order parameter $m_\mathrm{RE}$ (equation~\ref{eq.m}).
The heaviest REs Ho, Er and Tm disorder very quickly with temperature, losing
50\% of their ordering below 200~K.
By contrast, the Sm sublattice retains its ordering to much higher temperatures,
e.g.\ 50\% ordering at 650~K.
Although part of the reason for this behavior is the higher $\Tc$ of SmCo$_5$,
plots of the order parameter against reduced temperature $T/\Tc$ (not shown)
demonstrate that even when this factor is accounted for, Sm orders the most strongly.

Having an ordered RE at high temperature is useful for permanent magnets, since
the magnetocrystalline anisotropy decays faster than $m_\mathrm{RE}$~\cite{Kuzmin2008}.
Therefore SmCo$_5$ has a double advantage of having a high magnetocrystalline
anisotropy at low temperature, and a strong RE ordering to retain this anisotropy
at high temperature.

\subsection{Curie temperatures}

\begin{figure}
\includegraphics{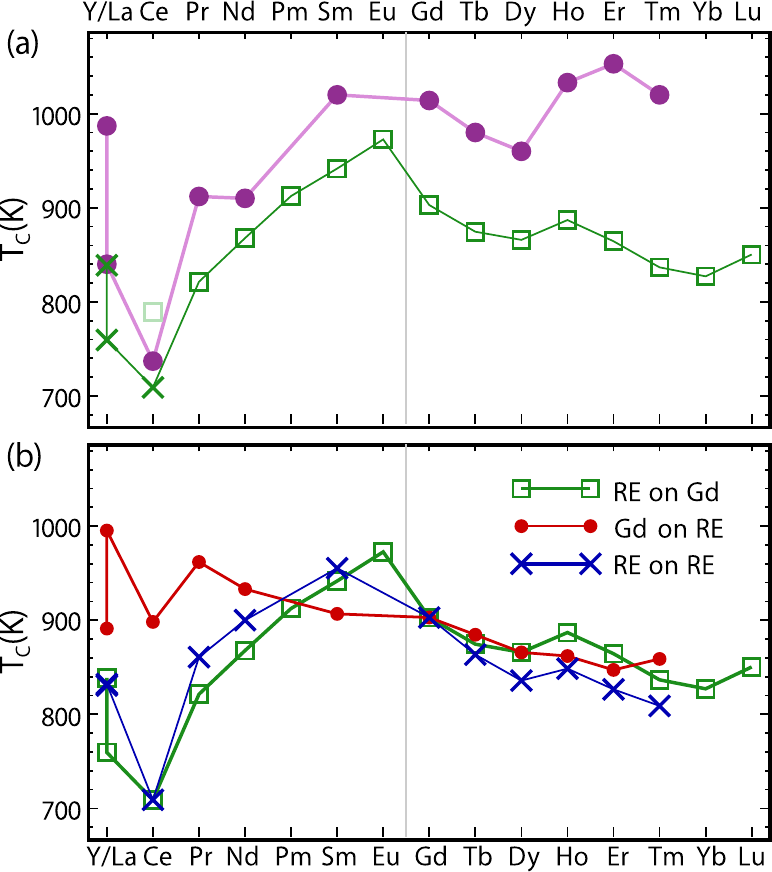}%
\caption{
(a) $\Tc$ calculated for RECo$_5$ using the GdCo$_5$
lattice parameters, compared to the experimental values
reported in Ref.~\cite{Buschow1977} (circles).
Squares and crosses are calculated with and without the LSIC respectively,
and the faint square is the energetically-unfavorable LSIC calculation
for CeCo$_5$.
(b) Comparison of $\Tc$ calculated for GdCo$_5$ using
RECo$_5$ lattice parameters (circles), RECo$_5$ using GdCo$_5$
lattice parameters (squares), and RECo$_5$ using RECo$_5$ lattice
parameters (crosses).
The faint gray lines separate light and heavy REs.
 \label{fig.Tc}}
\end{figure}

In Fig.~\ref{fig.Tc}(a) we compare the calculated Curie temperatures $\Tc$
(extracted from Fig.~\ref{fig.MvT}) to the experimental
values reported in Ref.~\cite{Buschow1977}.
We include $\Tc$ calculated for the non-forming Pm, Eu, Yb and LuCo$_5$
compounds.
We remind the reader that experimental values are for Co-rich heavy RECo$_5$
compounds, for which we would expect an increased $\Tc$.
For instance, the measured $\Tc$ of Gd$_2$Co$_{17}$ is 200~K higher 
than GdCo$_5$~\cite{Buschow1977}.

Fig.~\ref{fig.Tc}(a) clearly shows the contrasting
behavior of the light and heavy RECo$_5$ compounds.
Apart from YCo$_5$ and CeCo$_5$, the $\Tc$ of the 
light RECo$_5$ compounds increases monotonically
with the number of RE-4$f$ electrons.
Indeed, applying the energetically-unfavorable LSIC
to the Ce-4$f$ electron also causes CeCo$_5$ to follow
this trend [faint square in Fig.~\ref{fig.Tc}(a)].
Of the experimentally-known RECo$_5$ compounds,
SmCo$_5$ is calculated to have the highest $\Tc$ (942~K), 
but the $\Tc$ of the non-forming EuCo$_5$ compound is found to be 
even higher, at 973~K.

The trend in calculated $\Tc$ of the heavy RECo$_5$ compounds is less
obvious.
In general, filling the remaining subshell causes a reduction
in $\Tc$, but a secondary peak is observed at HoCo$_5$.
This peak in $\Tc$ coincides with a slight peak in Co moments for
HoCo$_5$ at zero temperature (Table.~\ref{tab.moments}).
Also, the $\Tc$ of LuCo$_5$ is very close to that calculated
for YCo$_5$ (850~and~839~K respectively).

The calculations and experiments agree on a number of qualititative features.
First, there is a substantial drop in $\Tc$ on moving from YCo$_5$
to LaCo$_5$, and another from LaCo$_5$ to CeCo$_5$.
As already noted, the drop for CeCo$_5$ is not observed
if the Ce-$4f$ electron is localized with the LSIC.
Second, SmCo$_5$ has the highest $\Tc$ of all the 
experimentally-attainable RECo$_5$ compounds.
Finally, the Co-rich heavy RECo$_5$ compounds 
do show a secondary peak in $\Tc$ like the calculations,
although at Er not Ho.
The heavy RE$_2$Co$_{17}$ compounds, whose stoichiometry
is better defined, also show a secondary peak around
Ho/Er/Tm followed by a sharp upturn for Lu~\cite{Buschow1977}.

The calculated variation in $\Tc$ shown in Fig.~\ref{fig.Tc}(a)
is only due to changing the RE.
In order to quantify the magnetostructural effect
of varying the lattice, we also calculated
$\Tc$ for the RECo$_5$ compounds using experimentally-reported 
lattice parameters (Table~\ref{tab.lattice}).
We further performed calculations where we varied 
the lattice but fixed the RE to Gd, i.e.\ GdCo$_5$
on different RECo$_5$ lattices.
We compare the three different sets of calculations in
Fig.~\ref{fig.Tc}(b).

First considering the calculations with the RE fixed
to Gd [red circles in Fig.~\ref{fig.Tc}(b)], 
we observe a decrease in $\Tc$ across the lanthanide block.
The exception is CeCo$_5$, which shows a strong 
magnetostructural effect;
as shown in Table~\ref{tab.lattice}, CeCo$_5$ has
an anomalously small $a$ parameter.
These calculations do not reproduce
experimental trends, e.g.\ predicting LaCo$_5$
to have the highest $\Tc$.

If instead we vary both the RE and the lattice parameters
[blue crosses in Fig.~\ref{fig.Tc}(b)] we find the
an almost identical trend in $\Tc$ as if we had kept
the lattice parameters fixed at GdCo$_5$ (green squares).
Using the RECo$_5$ lattice parameters accentuates
the drop in $\Tc$ for CeCo$_5$.
Unfortunately the experimentally-observed difference in $\Tc$
between YCo$_5$ and LaCo$_5$ is no longer calculated,
which can be seen as a cancellation of competing
green and red symbols in Fig.~\ref{fig.Tc}(b).
In general, the calculations find magnetostructural
effects to play a less important role in determining $\Tc$
than explicitly varying the RE.

\subsection{Order parameter expansion of the free energy}

Returning to the calculations with the lattice constants fixed to GdCo$_5$,
to gain further insight
into the calculated $\Tc$ we expand
the RDFT-DLM potential energy $\langle\Omega\rangle_{0,T}$ 
in terms of the order
parameters $m_\mathrm{RE}$, $m_\mathrm{Co_I}$ and $m_\mathrm{Co_{II}}$~\cite{Patrick2017}.
The labels I and II distinguish between the inequivalent Co positions in the RECo$_5$ structure
(Fig.~\ref{fig.RECo5}), i.e.\ the two Co atoms in plane with the RE 
(Co$_\mathrm{I}$, Wyckoff position $2c$) and the three out-of-plane 
Co atoms (Co$_\mathrm{II}$, Wyckoff position $3g$).
In this expansion, the Weiss field at each site
$(h_\mathrm{RE},h_\mathrm{Co_I},h_\mathrm{Co_{II}})$
is given by the equation
\begin{equation}
\begin{pmatrix}
h_\mathrm{RE} \\
2h_\mathrm{Co_I} \\
3h_\mathrm{Co_{II}}
\end{pmatrix}
=
\begin{pmatrix}
J_\mathrm{RE-RE}& J_{\mathrm{RE-Co_I}}& J_{\mathrm{RE-Co_{II}} }       \\
J_{\mathrm{RE-Co_I}}& J_{\mathrm{Co_I-Co_I}}& J_{\mathrm{Co_I-Co_{II}} }       \\
J_{\mathrm{RE-Co_{II}}}& J_{\mathrm{Co_{I}-Co_{II}}}& J_{\mathrm{Co_{II}-Co_{II}} }     
\end{pmatrix}
\begin{pmatrix}
m_\mathrm{RE} \\
m_\mathrm{Co_I} \\
m_\mathrm{Co_{II}}
\end{pmatrix}.
\label{eq.hmatrix}
\end{equation}
The prefactors in the Weiss fields account for the
site multiplicities.
The expansion of equation~\ref{eq.hmatrix} is 
valid for small $m$, i.e.\ close to $\Tc$.
The coefficients $J_{XY}$ are obtained by least-squares
fitting of RDFT-DLM calculations.
As discussed in Ref.~\cite{Patrick2017}, diagonalization
of the matrix in equation~\ref{eq.hmatrix} gives the RDFT-DLM
$\Tc$, thus allowing the variation shown in Fig.~\ref{fig.Tc}(a)
to be understood in terms of the strength of the interactions
between different magnetic sublattices.

\begin{figure}
\includegraphics{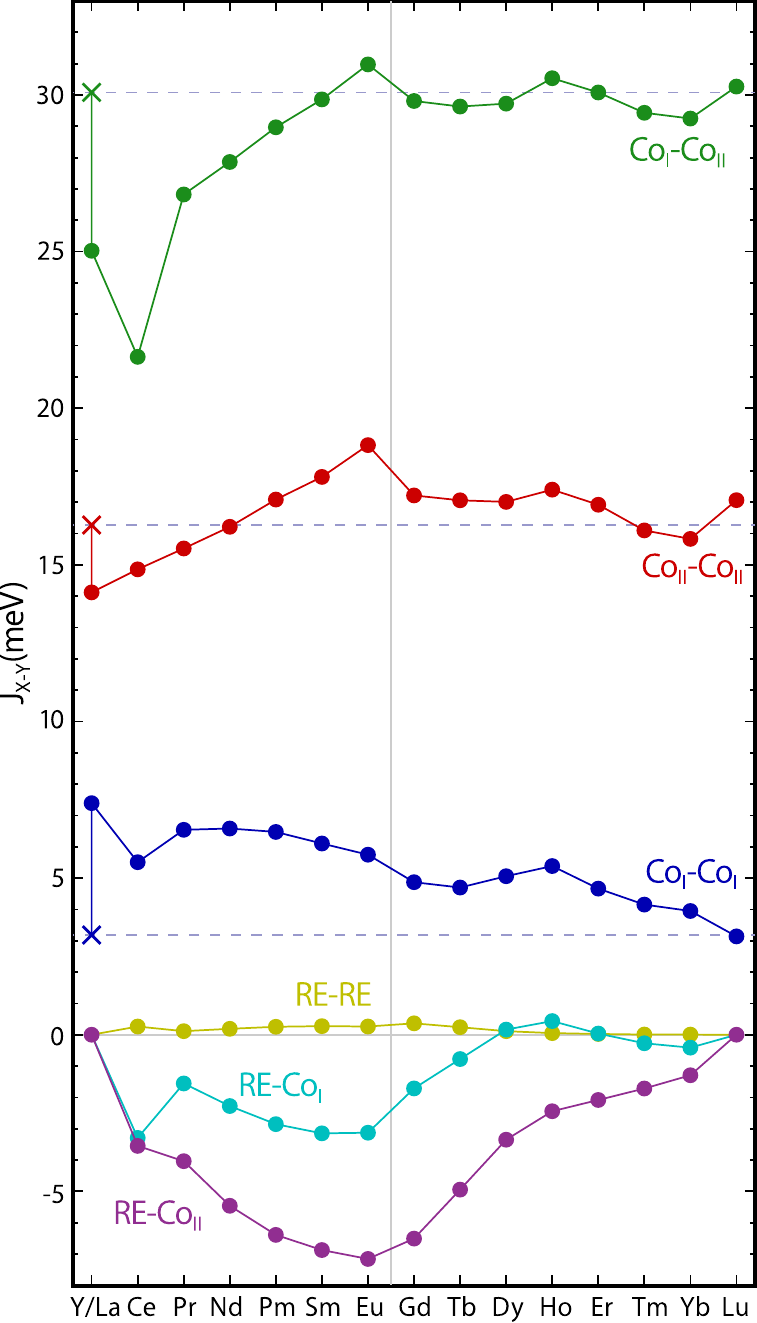}%
\caption{
Different $J_{XY}$ parameters (c.f.\ equation~\ref{eq.hmatrix})
calculated for RECo$_5$ on the GdCo$_5$ lattice 
Note that the Ce calculation was performed
without the LSIC, i.e. assuming that the Ce f-electron
is itinerant.
We highlight $J_{XY}$ for YCo$_5$ as crosses with 
horizontal dashed lines.
 \label{fig.Js}}
\end{figure}
The calculated coefficients $J_{XY}$ are shown in 
Fig.~\ref{fig.Js}.
A negative $J_{XY}$ indicates a tendency for species
$X$ and $Y$ to align antiferromagnetically.
Comparing Figs.~\ref{fig.Tc}(a) and \ref{fig.Js},
we see that the behavior of $\Tc$ is mirrored by
the largest $J_{XY}$ coefficient $J_{\mathrm{Co_I-Co_{II}}}$,
which describes the inter-layer Co interaction.
The next-largest coefficient $J_{\mathrm{Co_{II}-Co_{II}}}$,
describing the intra-layer interactions of the pure Co layer,
behaves similarly except that no drop at CeCo$_5$ is observed.
It is not surprising either that $\Tc$ tracks the largest 
$J_{XY}$ coefficients or that these coefficients describe
Co-Co interactions, in line with the picture that the TM
is responsible for the high $\Tc$ in RE-TM magnets.
What is less intuitive is that these coefficients should
be so strongly affected by the RE.

As found for $\Tc$, there is clear distinction between light
and heavy RECo$_5$ compounds for  $J_{\mathrm{Co_I-Co_{II}}}$
and $J_{\mathrm{Co_{II}-Co_{II}}}$.
By contrast $J_{\mathrm{Co_{I}-Co_{I}}}$ undergoes a general decrease
from La--Lu, with slight fluctuations around Ho and a dip at Ce.
The Co interactions are very similar for Y and Lu, consistent
with their similar $\Tc$.

The $J_{\mathrm{RE}-Y}$ coefficients which quantify RE
interactions are smaller in magnitude.
$J_{\mathrm{RE}-\mathrm{RE}}$ is particularly weak and
correlates with the size of the spin moment of the RE.
The strongest RE-Co interactions are interplanar,
RE-Co$_\mathrm{II}$.
Interestingly, neither $J_{\mathrm{Co_{II}-Co_{II}}}$
nor $J_{\mathrm{RE-Co_{II}}}$ show any strong anomaly
at CeCo$_5$, indicating that it is only the
Co$_\mathrm{I}$ interactions which are affected by the 
itinerant Ce-$4f$ electron.

Again comparing the light and heavy REs,
we note that the in-plane interaction quantified
by $J_{\mathrm{RE-Co_{I}}}$ actually becomes ferromagnetic
for DyCo$_5$, HoCo$_5$ and ErCo$_5$, which coincides with the secondary peak in $\Tc$
[Fig.~\ref{fig.Tc}(a)].
Also, we observe that the strongest RE-Co interactions
occur not for GdCo$_5$, which has the largest RE spin moment,
but rather EuCo$_5$.
\section{Discussion}
\label{sec.discussion}

\subsection{The RE-TM interaction}
\begin{figure}
\includegraphics{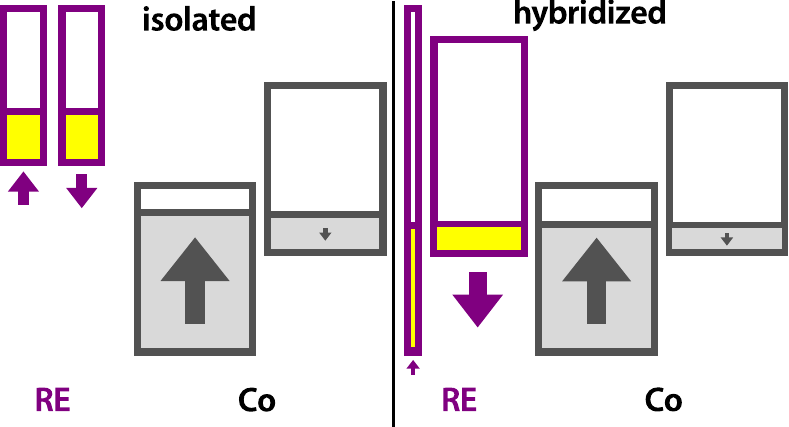}%
\caption{Schematic of antiferromagnetic RE-Co interaction,
after Fig.~2 of Ref.~\cite{Brooks1989}. 
Wide/narrow rectangles symbolize strong/weak RE-Co 
hybridization in a given spin channel.
\label{fig.RETM}}
\end{figure}

Our calculations have found that the strength of magnetic
interactions between Co moments in RECo$_5$ is affected by the
RE, even when the lattice parameters are held fixed.
As a result of this variation, $\Tc$ depends heavily
on the RE.
To explain this behavior, we first recall the theory of 
RE-TM interactions described in Ref.~\cite{Brooks1989},
which explains the antiferromagnetic spin coupling in
terms of hybridization between RE-5$d$ and TM-3$d$ states.
Figure~\ref{fig.RETM} is a schematic representation of the theory.
The magnetic properties of Co originate from almost-full
and almost-empty $3d$ bands of opposing spins.
The minority spin Co-3$d$ band lies closer in energy to the
RE-5$d$ bands than the majority Co-$d$ band, and therefore
hybridizes more strongly.
The preferential occupation of the lower-energy hybridized
spin states causes the RE-5$d$ bands to develop a spin
polarization in the direction of the Co minority spin,
i.e.\ an antiferromagnetic coupling.
Then, onsite RE 4$f$-5$d$ interactions polarize the RE-4$f$
spins in the same sense.

In this picture, the RE-TM interaction
varies according to the strength of the 4$f$-5$d$
interaction, which is expected to scale with the spin
moment of the RE.
Accordingly, the strongest RE-TM interactions are expected for 
Gd.
However, any effects on the TM magnetization are expected
to proceed via the Co-3$d$-RE-5$d$ hybridization, with
no direct link to the RE-$4f$ states.

\subsection{Magnetostructural effects}

Within the picture of Fig.~\ref{fig.RETM}, any 
variation in $\Tc$ implies that 
the RE-5$d$ states are not the same for all REs.
Of course, the RE-5$d$ orbitals do vary across
the lanthanide block in terms of their spatial extent,
as can be seen from the experimental
lattice parameters in Table~\ref{tab.lattice}.
The lattice parameter $a$ of LaCo$_5$ is
3\% larger than GdCo$_5$, while for YCo$_5$ the
difference is less than 0.5\%.
The experimental lattice parameters of LuCo$_5$ are not
known, but the ionic radius of Lu is much closer to Y
than La~\cite{Taylorbook}.
Correspondingly, the $\Tc$ values calculated at 
GdCo$_5$ lattice parameters are much closer
for YCo$_5$ and LuCo$_5$ (11~K) than
YCo$_5$ and LaCo$_5$ (80~K).

So, independent of any arguments based on the RE-4$f$
states, the calculations on YCo$_5$, LaCo$_5$
and LuCo$_5$ suggest that the size of the 
RE-5$d$ orbitals affects the Co magnetism.
Indeed we could have reached a similar conclusion
from our calculations on GdCo$_5$ with variable
lattice parameter.
Using the lattice parameters of lighter (heavier) RECo$_5$ compounds
for GdCo$_5$ corresponds to expansion (compression) of $a$
(Table~\ref{tab.lattice}).
From the red line of Fig.~\ref{fig.Tc}(b), we see that expansion
of $a$ is correlated with an increased $\Tc$, while compression
reduces it.
Inversely, using GdCo$_5$ lattice parameters for the light 
and heavy RECo$_5$ compounds corresponds
to compression and expansion of $a$  respectively.
Comparing the green and blue symbols in Fig.~\ref{fig.Tc}(b)
confirms that compression reduces $\Tc$ (green lower than blue
for La--Gd) while expansion increases $\Tc$ (green
higher than blue for Gd--Tm).

This magnetostructural effect makes some contribution to
the overall variation of $\Tc$.
Interestingly, the coefficients in Fig.~\ref{fig.Js}
which quantify the Co$_\mathrm{I}$-Co$_\mathrm{I}$ interaction
(blue symbols) resemble the 
behavior of $\Tc$ calculated for GdCo$_5$ with different
lattice parameters [red symbols in Fig.~\ref{fig.Tc}(b)].
Taken together with the fact that these Co$_\mathrm{I}$
atoms sit in plane with the RE atoms (Fig.~\ref{fig.RECo5}),
we assert that the variation $J_{\mathrm{Co_I-Co_{I}}}$
is magnetostructural in origin,
with the RE-5$d$ orbitals affecting the in-plane Co-3$d$
interactions.

However, magnetostructural effects cannot
really explain the observed
variation in $\Tc$.
First, they do not account for the qualitative difference
in behavior between light and heavy RECo$_5$ compounds.
Second, the $J_{\mathrm{Co_I-Co_{I}}}$ coefficients which
are sensitive to the structure do not play a major
role in determining $\Tc$, compared to $J_{\mathrm{Co_I-Co_{II}}}$ 
and $J_{\mathrm{Co_{II}-Co_{II}}}$.
For example, LaCo$_5$ has the largest  $J_{\mathrm{Co_I-Co_{I}}}$
but the second lowest $\Tc$ [Fig.~\ref{fig.Tc}(a)].
Therefore, we look for an additional explanation.

\subsection{Densities-of-states}

\begin{figure}
\includegraphics{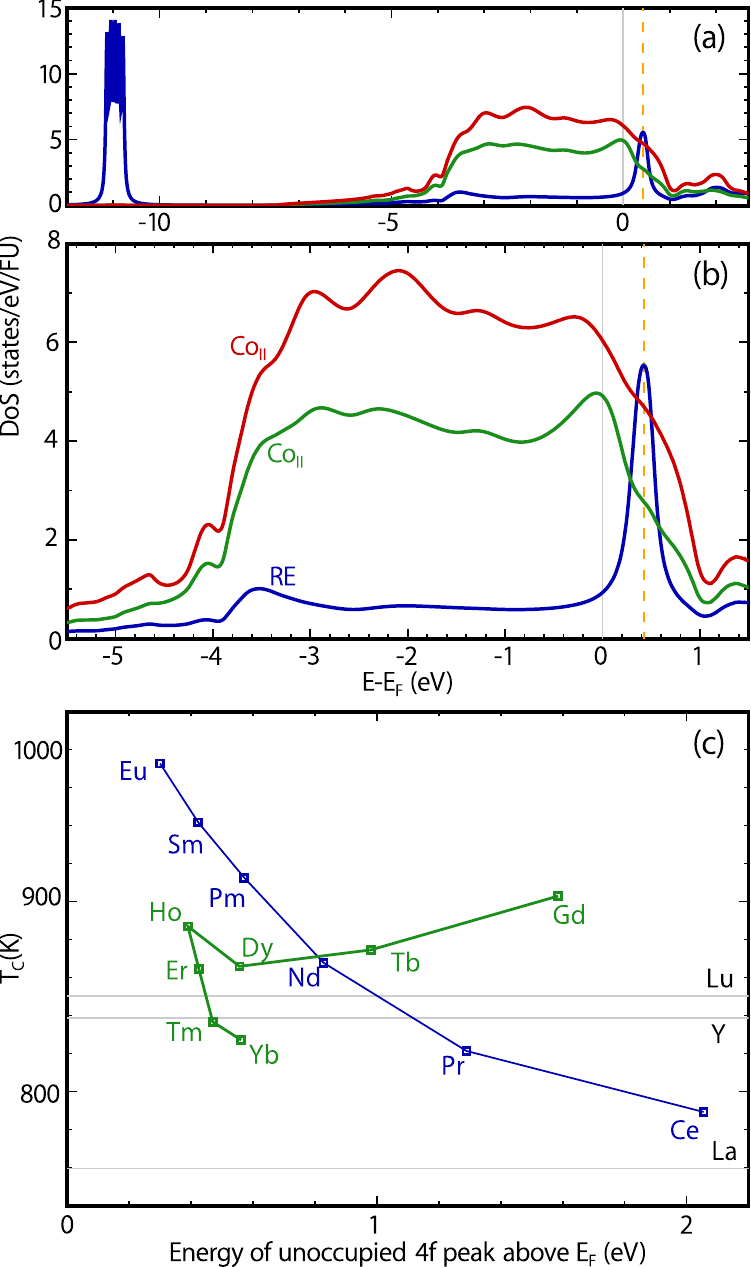}%
\caption{%
DoS calculated just below $\Tc$
for SmCo$_5$, resolved onto the Sm, Co$_\mathrm{I}$ and Co$_\mathrm{II}$
sublattices, shown (a) across a wide energy scale and (b) around the Fermi energy.
The vertical dashed line intersects the center of the unoccupied
4$f$ peak above the Fermi energy.
(c) $\Tc$ plotted against the center of this
unoccupied peak for the different RECo$_5$ compounds.
Note that here the value of $\Tc$ of
CeCo$_5$ was calculated with the LSIC applied.
\label{fig.DoS}}
\end{figure}

In Fig.~\ref{fig.DoS}(a) we plot the DFT Kohn-Sham density-of-states (DoS)
of SmCo$_5$.
The DoS was calculated just below $\Tc$ (i.e.\ at an almost
completely disordered state), using GdCo$_5$ lattice parameters,
and has been resolved into contributions from the RE, Co$_\mathrm{I}$
and Co$_\mathrm{II}$ sublattices.
The wide energy scale of Fig.~\ref{fig.DoS}(a) was chosen
to show explicitly the energy position of the occupied Sm-4$f$
states, 11~eV below the Fermi level $E_F$.
Zooming in on the region around $E_F$ [Fig.~\ref{fig.DoS}(b)]
shows the Co-$3d$ band (bandwidth $\sim$4~eV) hybridized
with the RE-$5d$ states.
However, an additional prominent feature is observed
in the RE DoS, which is a narrow peak above $E_F$.
The weight of this peak is approximately two electrons,
and corresponds to the two unoccupied RE-$4f$ 
states in the $\downarrow$ spin channel (Fig.~\ref{fig.Hund}).

A similar peak can be observed in the DoS of
all the RECo$_5$ compounds.
For REs with almost empty 4$f$ spin subshells, the peak
is located at high energy, and comes closer to $E_F$
as the subshell becomes filled (for light REs, 
a second peak corresponding to the opposite
spin channel is also present, at much higher energies).
We stress that in our DFT description, states
above $E_F$ make no contribution to calculated
properties.
However, the tail of this unoccupied RE-4$f$ peak does extend
below $E_F$ and therefore contributes to the density.
In fact, this tail is the origin of the noninteger contribution
to the $f$-resolved spin moments pointed out in Sec.~\ref{sec.decomp}
when discussing Table~\ref{tab.moments}.

As indicated in Fig.~\ref{fig.DoS}(b), we can extract
the energy corresponding to the centre of this unoccupied peak (dashed line).
Then, in Fig.~\ref{fig.DoS}(c) we plot the calculated $\Tc$
as a function of this peak position.
The light REs show an apparently
strong correlation, with $\Tc$ increasing as the
unoccupied peak becomes closer to $E_F$.
By contrast the heavy REs do not show any particular
correlation.
The possible exception is HoCo$_5$, which as well
as having a higher $\Tc$ than
its neighbors also has the unoccupied RE-4$f$ peak closest
to $E_F$.

An explanation for the differing behavior of the light
and heavy RECo$_5$ compounds in Fig.~\ref{fig.DoS}(c)
relates to the spin character of the unoccupied
peak.
For the light REs, the unoccupied RE-4$f$ peak closest
to $E_F$ has the same $\downarrow$ spin as the Co-3$d$ minority spins,
i.e.\ the states which hybridize strongly with
the RE-5$d$ states and lead to antiferromagnetic coupling 
(Fig.~\ref{fig.RETM}).
By contrast, the unoccupied RE-4$f$ peak of the heavy
REs has the same $\uparrow$ spin character as the Co-$d$ majority
spins.
The hybridization of these states with RE-5$d$ is weak
due to the energy separation; also, it favors ferromagnetic
coupling.
As noted when discussing Fig.~\ref{fig.Js}, HoCo$_5$
does indeed have a positive $J_{\mathrm{RE}-\mathrm{Co_I}}$
coefficient, corresponding to a ferromagnetic RE-TM interaction.
Indeed the temperature evolution of the order parameters
in Fig.~\ref{fig.orderparams}
shows how the overall antiferromagnetic RE-TM coupling
is weakened for the heavy RECo$_5$ compounds.

We therefore propose a mechanism
where a small contribution of $f$-character 
RE states, located just below the Fermi level,
affects $\Tc$ by modifying the Co-3$d$ states,
probably indirectly through the RE-5$d$ states.
Such a mechanism could explain why we calculate
higher $\Tc$s than GdCo$_5$ for Pm, Sm, and EuCo$_5$, 
despite these elements having smaller
spin moments and being placed on a lattice
with a compressed $a$ parameter.
The effect is strong (weak) for the light (heavy) 
RECo$_5$ compounds, and favors antiferromagnetic
(ferromagnetic) RE-TM coupling as described above,
consistent with the behavior of $J_{\mathrm{RE}-\mathrm{Co_I}}$
and  $J_{\mathrm{RE}-\mathrm{Co_{II}}}$ shown
in Fig.~\ref{fig.Js}.

We have already pointed out that the calculations
have found SmCo$_5$ both to have the highest $\Tc$
of the experimentally-realized RECo$_5$ compounds
and also a strong RE-TM interaction, which enables Sm 
to stay ordered at high temperature.
Within the mechanism described here, the origin of this
behavior is Sm's almost-filled 4$f$ spin subshell.
The hypothetical EuCo$_5$ compound would have an even higher $\Tc$,
but unfortunately does not form.
The total energies calculated at the scalar-relativistic level
find Eu to be more stable in the 3+ state than 2+, when
forced to occupy the RECo$_5$ structure.
However, we have not investigated the full compositional phase
diagram where different stoichiometries and structures might
have a lower free energy.

\section{Outlook}
\label{sec.conclusion}

We have devised a physically transparent theory to model the magnetic
properties of RE-TM compounds, with particular emphasis
on their finite temperature properties.
The magnetic disorder is described with the disordered local moment
picture based on relativistic density-functional theory, with the 
RE-4$f$ electrons treated with the local self-interaction correction
which encapsulates Hund's rules.
We used the theory to calculate 
the zero and finite temperature
properties of the RECo$_5$ family of magnets, comparing
magnetic moments and Curie temperatures to experimental
measurements.

When presenting our theory we stated that, mainly as a result of
the spherical approximations and mean-field nature of the theory, 
we expected our approach to perform best in calculating trends across
a series.
This statement has been borne out by our comparisons with experimental
data, where we were able to reproduce a number of qualitative features.
In particular we were able to track the behavior of $\Tc$, which to
our knowledge has never been accomplished from first principles
before.

We identified interesting behavior from the calculations, that even
though $\Tc$ is generally determined by TM-TM interactions, these
interactions were nonetheless affected by the RE.
We argued that while the varying spatial extent of the RE-5$d$ orbitals
did affect the TM-TM interactions, this effect was not sufficient
to explain the variation in $\Tc$.
Instead, we proposed a mechanism based on a small $f$-character
contribution to the density around the Fermi level which,
for the light RECo$_5$ compounds, strengthens both the
RE-TM and TM-TM interactions.

We note that more expensive DMFT calculations do not provide
an obvious pathway for a further exploration of this mechanism,
neither in being able to calculate $\Tc$, nor also 
since we cannot make any assumptions
about the hybridization of the RE-4$f$ electrons~\cite{Delange2017}.
In terms of experimental evidence, we currently have only
the observation that SmCo$_5$ has a higher $\Tc$
than GdCo$_5$.
To our knowledge, this observation has not been
explained before, but on its own cannot be considered
justification for the correctness of the LSIC.
However, the theory presented here opens the door to 
performing a more detailed comparision with experimental
measurements on the temperature-dependent properties of 
any RE-TM compound,
as was already done for YCo$_5$ and GdCo$_5$~\cite{Patrick2017}.

Beyond exploring the fundamental 
physics of RE-TM magnets, our theoretical framework allows
the study of practical aspects.
In particular, the CPA formalism allows the effects 
of compositional disorder, e.g.\ substitution of RE
or TM elements, to be investigated.
Furthermore, with a view to optimizing high-temperature coercivity,
it is highly desirable to tackle the temperature
dependence of the magnetocrystalline anisotropy~\cite{Staunton2004}.
Such calculations require a careful incorporation of crystal-field
effects into our ASA calculations~\cite{Hummler1996} and
also special care regarding how the calculated quantities are
compared to experimental measurements, given the ferrimagnetic
nature of the RE-TM magnets~\cite{Patrick2018}.
Already the current calculations have found the high-temperature 
RE magnetic ordering to be strongest in SmCo$_5$, the best-performing
magnet in the RECo$_5$ family.

\appendix
\section{Coupling introduced by the LSIC}
\label{app.coupling}
Here we list the formulae for the different coupling functions
which enter the coupled equations~\ref{eq.coupled}.
Again we emphasize that $\kappa = -l-1$ and $\kappa' = l$.
We have also introduced the quantities $\bar{l}_1 = l+1$ and $\bar{l}_2 = l - 1$.
\begin{widetext}
\begin{eqnarray}
\mathcal{G}^{m_j}_\pm(\kappa,\kappa) &=& \frac{2 m_j}{2l+1} B_\mathrm{XC}
\pm\left[
V_{l(m_j-1/2)}^{\uparrow}
\left(\frac{l + m_j + 1/2}{2l+1}\right) 
+ V_{l(m_j+1/2)}^{\downarrow}
\left(\frac{l - m_j + 1/2}{2l+1}\right)
\right]
 \nonumber \\
\mathcal{G}^{m_j}_\pm(\kappa',\kappa') &=& -\frac{2 m_j}{2l+1} B_\mathrm{XC}
\pm\left[
V_{l(m_j-1/2)}^{\uparrow}
\left(\frac{l - m_j + 1/2}{2l+1}\right) 
+ V_{l(m_j+1/2)}^{\downarrow}
\left(\frac{l + m_j + 1/2}{2l+1}\right)
\right]
 \nonumber \\
\mathcal{G}^{m_j}_\pm(\kappa,\kappa') &=& 
-\left( 1 - \frac{m_j^2}{(l+1/2)^2}\right)^{\frac{1}{2}} \left[ B_\mathrm{XC}
\mp\frac{V_{l(m_j-1/2)}^{\uparrow} - V_{l(m_j+1/2)}^{\downarrow}}{2}\right]
= \mathcal{G}^{m_j}_\pm(\kappa',\kappa) \nonumber \\
\mathcal{G}^{m_j}_\pm(-\kappa,-\kappa) &=& -\frac{2 m_j}{2\bar{l}_1+1} B_\mathrm{XC}
\pm\left[
V_{\bar{l}_1(m_j-1/2)}^{\uparrow}
\left(\frac{\bar{l}_1 - m_j + 1/2}{2\bar{l}_1+1}\right) 
+ V_{\bar{l}_1(m_j+1/2)}^{\downarrow}
\left(\frac{\bar{l}_1 + m_j + 1/2}{2\bar{l}_1+1}\right)
\right]
 \nonumber \\
\mathcal{G}^{m_j}_\pm(-\kappa',-\kappa') &=& -\frac{2 m_j}{2\bar{l}_2+1} B_\mathrm{XC}
\pm\left[
V_{\bar{l}_2(m_j-1/2)}^{\uparrow}
\left(\frac{\bar{l}_2 - m_j + 1/2}{2\bar{l}_2+1}\right) 
+ V_{\bar{l}_2(m_j+1/2)}^{\downarrow}
\left(\frac{\bar{l}_2 + m_j + 1/2}{2\bar{l}_2+1}\right)
\right]
 \nonumber
\end{eqnarray} 
\end{widetext}

\section{Relativistic couplings between different spin-orbital channels}
\label{app.mixing}
\begin{figure*}
\includegraphics{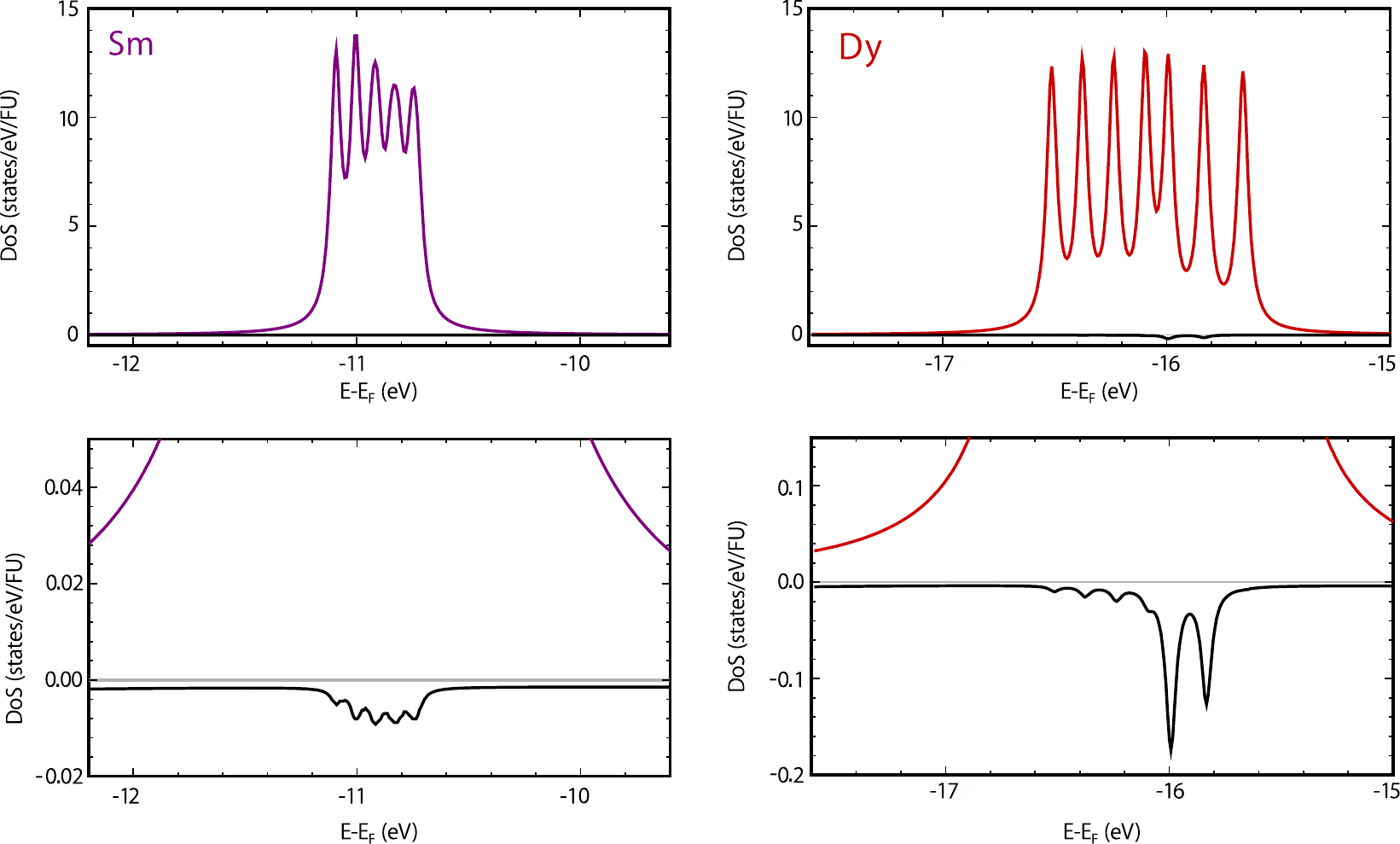}%
\caption{
Spin-resolved, zero-temperature DoS for SmCo$_5$ (left) and DyCo$_5$ (right),
at energies corresponding to the majority-spin, SI-corrected states.
The majority and minority spin contributions are plotted with positive
and negative signs, respectively, and the lower plots zoom in on the minority
contribution.
Note the larger scale for DyCo$_5$.
\label{fig.mixing}
}
\end{figure*}
As indicated by equation~\ref{eq.mixing},
the spin-orbit interaction mixes different ($\sigma,m$) channels, 
including those which do and do not have the LSIC applied.
In general, since there is a large energy separation between corrected 
and uncorrected states ($\sim$10~eV), the energy denominator that 
appears in the perturbative expansion of the state is large and thus
the mixing is small.
Nonetheless, the mixing can be seen by examining
the zero-temperature, spin-resolved DoS at energies
around the occupied (majority spin) 4$f$ electrons.

This quantity is plotted in Fig.~\ref{fig.mixing} for 
SmCo$_5$ and DyCo$_5$.
In the scalar-relativistic calculation the occupied 4$f$ electrons
are spin pure, but on performing the relativistic calculation
a small contribution appears in the minority spin channel (negative
scale in Fig.~\ref{fig.mixing}), due to the mixing described above.
This contribution is bigger for DyCo$_5$ than SmCo$_5$ (note change of scale) 
because there are two SI-corrected minority spin states located
4~eV above the majority spin peak which mix more strongly.
For SmCo$_5$ the mixing only occurs with SI-uncorrected states lying above
the Fermi level.
The large energy separation suppresses the mixing in this case.

\section{Zero-temperature moments calculated at experimental lattice parameters}
\label{app.explatt}

\begin{figure}
\includegraphics{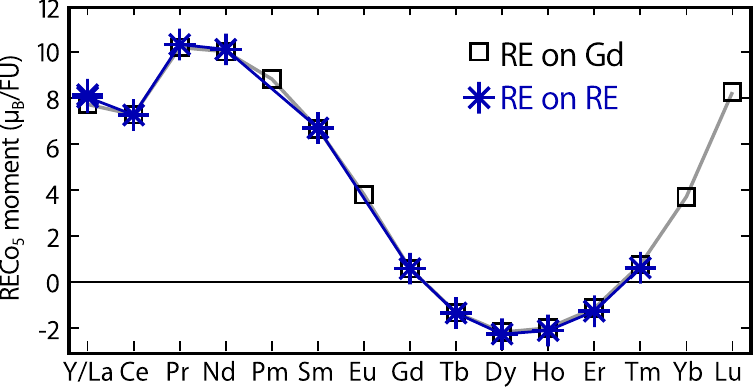}%
\caption{
Magnetic moments calculated at zero temperature for 
RECo$_5$ using GdCo$_5$
lattice parameters (squares), and RECo$_5$ using RECo$_5$ lattice
parameters (stars).
\label{fig.moms_explatt}}
\end{figure}

In Fig.~\ref{fig.moms_explatt} we compare the zero temperature
moments calculated either using GdCo$_5$ lattice parameters or,
where available, RECo$_5$ lattice parameters (Table~\ref{tab.lattice}).
Note that these calculations were performed without the OPC applied,
which results in reduced Co moments compared to Fig.~\ref{fig.zeroTmoms}.

\begin{acknowledgments}
The present work forms part of the PRETAMAG project, 
funded by the UK Engineering and  Physical  Sciences 
Research  Council (EPSRC),  Grant  no.\   EP/M028941/1.
\end{acknowledgments}
%
\end{document}